\PassOptionsToPackage{unicode}{hyperref}
\PassOptionsToPackage{hyphens}{url}
\PassOptionsToPackage{dvipsnames,svgnames,x11names}{xcolor}
\documentclass[
  11pt,
  a4paper,
]{article}

\usepackage{amsmath,amssymb}
\usepackage{setspace}
\usepackage{iftex}
\ifPDFTeX
  \usepackage[T1]{fontenc}
  \usepackage[utf8]{inputenc}
  \usepackage{textcomp} 
\else 
  \usepackage{unicode-math}
  \defaultfontfeatures{Scale=MatchLowercase}
  \defaultfontfeatures[\rmfamily]{Ligatures=TeX,Scale=1}
\fi
\usepackage{lmodern}
\ifPDFTeX\else  
\fi
\IfFileExists{upquote.sty}{\usepackage{upquote}}{}
\IfFileExists{microtype.sty}{
  \usepackage[]{microtype}
  \UseMicrotypeSet[protrusion]{basicmath} 
}{}
\makeatletter
\@ifundefined{KOMAClassName}{
  \IfFileExists{parskip.sty}{%
    \usepackage{parskip}
  }{
    \setlength{\parindent}{0pt}
    \setlength{\parskip}{6pt plus 2pt minus 1pt}}
}{
  \KOMAoptions{parskip=half}}
\makeatother
\usepackage{xcolor}
\usepackage[top=2.5cm,bottom=2.5cm,left=2.5cm,right=2.5cm]{geometry}
\providecommand{\tightlist}{%
  \setlength{\itemsep}{0pt}\setlength{\parskip}{0pt}}\usepackage{longtable,booktabs,array}
\usepackage{calc} 
\usepackage{etoolbox}
\makeatletter
\patchcmd\longtable{\par}{\if@noskipsec\mbox{}\fi\par}{}{}
\makeatother
\IfFileExists{footnotehyper.sty}{\usepackage{footnotehyper}}{\usepackage{footnote}}
\makesavenoteenv{longtable}
\usepackage{graphicx}
\makeatletter
\def\maxwidth{\ifdim\Gin@nat@width>\linewidth\linewidth\else\Gin@nat@width\fi}
\def\maxheight{\ifdim\Gin@nat@height>\textheight\textheight\else\Gin@nat@height\fi}
\makeatother
\setkeys{Gin}{width=\maxwidth,height=\maxheight,keepaspectratio}
\makeatletter
\def\fps@figure{htbp}
\makeatother

\usepackage{booktabs}
\usepackage{longtable}
\usepackage{array}
\usepackage{multirow}
\usepackage{wrapfig}
\usepackage{float}
\usepackage{colortbl}
\usepackage{pdflscape}
\usepackage{tabu}
\usepackage{threeparttable}
\usepackage{threeparttablex}
\usepackage[normalem]{ulem}
\usepackage{makecell}
\usepackage{xcolor}
\usepackage{orcidlink}
\definecolor{mypink}{RGB}{219, 48, 122}
\usepackage{todonotes}
\usepackage[bibliography=common]{apxproof}
\def\Var{\operatorname{Var}}
\def\E{\operatorname{E}}
\def\tr{\operatorname{tr}}
\makeatletter
\@ifpackageloaded{caption}{}{\usepackage{caption}}
\AtBeginDocument{%
\ifdefined\contentsname
  \renewcommand*\contentsname{Table of contents}
\else
  \newcommand\contentsname{Table of contents}
\fi
\ifdefined\listfigurename
  \renewcommand*\listfigurename{List of Figures}
\else
  \newcommand\listfigurename{List of Figures}
\fi
\ifdefined\listtablename
  \renewcommand*\listtablename{List of Tables}
\else
  \newcommand\listtablename{List of Tables}
\fi
\ifdefined\figurename
  \renewcommand*\figurename{Figure}
\else
  \newcommand\figurename{Figure}
\fi
\ifdefined\tablename
  \renewcommand*\tablename{Table}
\else
  \newcommand\tablename{Table}
\fi
}
\@ifpackageloaded{float}{}{\usepackage{float}}
\floatstyle{ruled}
\@ifundefined{c@chapter}{\newfloat{codelisting}{h}{lop}}{\newfloat{codelisting}{h}{lop}[chapter]}
\floatname{codelisting}{Listing}

\usepackage{amsthm}
\theoremstyle{definition}
\newtheorem{example}{Example}[section]
\theoremstyle{plain}
\newtheorem{theorem}{Theorem}[section]
\theoremstyle{plain}
\newtheorem{corollary}{Corollary}[section]
\theoremstyle{plain}
\newtheorem{lemma}{Lemma}[section]
\theoremstyle{remark}
\AtBeginDocument{}

\makeatother
\makeatletter
\@ifpackageloaded{caption}{}{\usepackage{caption}}
\@ifpackageloaded{subcaption}{}{\usepackage{subcaption}}
\makeatother
\ifLuaTeX
  \usepackage{selnolig}  
\fi
\usepackage[style=authoryear-comp,]{biblatex}
\usepackage{bookmark}

\IfFileExists{xurl.sty}{\usepackage{xurl}}{} 
\hypersetup{
  pdftitle={Forecast Linear Augmented Projection (FLAP): A free lunch to reduce forecast error variance},
  pdfauthor={Yangzhuoran Fin Yang; George Athanasopoulos; Rob~J Hyndman; Anastasios Panagiotelis},
  colorlinks=true,
  linkcolor={blue},
  filecolor={Maroon},
  citecolor={Blue},
  urlcolor={Blue},
  pdfcreator={LaTeX via pandoc}}

\usepackage{caption}
\DeclareCaptionStyle{italic}[justification=centering]
 {labelfont={bf},textfont={it},labelsep=colon}
\usepackage{bera}
\usepackage[charter]{mathdesign}
\usepackage[scale=0.9]{sourcecodepro}
\usepackage[lf,t]{FiraSans}

\usepackage{fancyhdr}
\fancypagestyle{plain}{%
\fancyhf{} 
\fancyfoot[C]{\sffamily\thepage} 

}

\usepackage{bm,amsmath}
\allowdisplaybreaks

\makeatletter
\def\fps@figure{htbp}
\makeatother
\usepackage[compact,sf,bf]{titlesec}
\titleformat{\section}[block]
  {\fontsize{15}{17}\bfseries\sffamily}
  {\thesection}
  {0.4em}{}
\titleformat{\subsection}[block]
  {\fontsize{12}{14}\bfseries\sffamily}
  {\thesubsection}
  {0.4em}{}
\titlespacing{\section}{0pt}{*5}{*1}
\titlespacing{\subsection}{0pt}{*2}{*0.2}
\titlespacing{\subsubsection}{0pt}{*1}{*0.1}

\makeatletter
\@ifpackageloaded{biblatex}{
\ExecuteBibliographyOptions{bibencoding=utf8,minnames=1,maxnames=3, maxbibnames=99,dashed=false,terseinits=true,giveninits=true,uniquename=false,uniquelist=false,doi=false, isbn=false,url=true,sortcites=false}
\DeclareFieldFormat{url}{\texttt{\url{#1}}}
\DeclareFieldFormat[article]{pages}{#1}
\DeclareFieldFormat[inproceedings]{pages}{\lowercase{pp.}#1}
\DeclareFieldFormat[incollection]{pages}{\lowercase{pp.}#1}
\DeclareFieldFormat[article]{volume}{\mkbibbold{#1}}
\DeclareFieldFormat[article]{number}{\mkbibparens{#1}}
\DeclareFieldFormat[article]{title}{\MakeCapital{#1}}
\DeclareFieldFormat[article]{url}{}
\DeclareFieldFormat[inproceedings]{title}{#1}
\DeclareFieldFormat{shorthandwidth}{#1}
\usepackage{xpatch}
\xpatchbibmacro{volume+number+eid}{\setunit*{\adddot}}{}{}{}
\renewbibmacro{in:}{%
  \ifentrytype{article}{}{%
  \printtext{\bibstring{in}\intitlepunct}}}
\AtEveryBibitem{\clearfield{month}}
\AtEveryCitekey{\clearfield{month}}
\DeclareDelimFormat[cbx@textcite]{nameyeardelim}{\addspace}

}{}
\makeatother

\usepackage{microtype}

\RequirePackage[absolute,overlay]{textpos}
\def\placefig#1#2#3#4{\begin{textblock}{.1}(#1,#2)\rlap{\includegraphics[#3]{#4}}\end{textblock}}

\title{Forecast Linear Augmented Projection (FLAP): A free lunch to
reduce forecast error variance}
\date{2 July 2024}

\def\Date{\number\day}
\def\Month{\ifcase\month\or
 January\or February\or March\or April\or May\or June\or
 July\or August\or September\or October\or November\or December\fi}
\def\Year{\number\year}

\makeatletter
\def\wp#1{\gdef\@wp{#1}}\def\@wp{??/??}
\def\jel#1{\gdef\@jel{#1}}\def\@jel{??}
\def\showjel{{\large\textsf{\textbf{JEL classification:}}~\@jel}}
\def\nojel{\def\showjel{}}
\makeatother

\nojel

\makeatletter
\def\cover{{\sffamily\setcounter{page}{0}
        \thispagestyle{empty}
        \placefig{2}{1.5}{width=5cm}{monash2}
        \placefig{16.9}{1.5}{width=2.1cm}{MBSportrait}
        \begin{textblock}{4}(16.9,4)ISSN 1440-771X\end{textblock}
        \begin{textblock}{7}(12.7,27.9)\hfill
        \includegraphics[height=0.7cm]{AACSB}~~~
        \includegraphics[height=0.7cm]{EQUIS}~~~
        \includegraphics[height=0.7cm]{AMBA}
        \end{textblock}
        \vspace*{2.5cm}
        \begin{center}\Large
        Department of Econometrics and Business Statistics\\[.5cm]
        \footnotesize http://monash.edu/business/ebs/research/publications
        \end{center}\vspace{2cm}
        \begin{center}
        \fbox{\parbox{14cm}{\begin{onehalfspace}\centering\Huge\vspace*{0.3cm}
                \textsf{\textbf{\expandafter{\@title}}}\vspace{1cm}\par
                \LARGE
                \expandafter{\@author}
                \end{onehalfspace}
        }}
        \end{center}
        \vfill
                \begin{center}\Large
                \Month~\Year\\[1cm]
                Working Paper \@wp
        \end{center}\vspace*{2cm}}}
        \def\addresses#1{\gdef\@addresses{#1}}\def\@addresses{??}
        \def\pageone{{\sffamily\setstretch{1}%
        \thispagestyle{empty}%
        \vbox to \textheight{%
        \raggedright\baselineskip=1.2cm
     {\fontsize{24.88}{30}\sffamily\textbf{\expandafter{\@title}}}
        \vspace{2cm}\par
        \hspace{1cm}\parbox{14cm}{\sffamily\large\@addresses}\vspace{1cm}\vfill
        \hspace{1cm}{\large\Date~\Month~\Year}\\[1cm]
        \hspace{1cm}\showjel\vss}}}
\def\blindtitle{{\sffamily
     \thispagestyle{plain}\raggedright\baselineskip=1.2cm
     {\fontsize{24.88}{30}\sffamily\textbf{\expandafter{\@title}}}\vspace{1cm}\par
        }}
\def\titlepage{{\cover\newpage\pageone\newpage\blindtitle}}

\def\blind{\def\titlepage{{\blindtitle}}\let\maketitle\blindtitle}
\def\titlepageonly{\def\titlepage{{\pageone\end{document}}}}
\def\nocover{\def\titlepage{{\pageone\newpage\blindtitle}}\let\maketitle\titlepage}
\let\maketitle\titlepage
\makeatother

\nocover
  \author{Yangzhuoran Fin Yang, George Athanasopoulos, Rob~J
Hyndman, Anastasios Panagiotelis}
  \addresses{%
      \textbf{Yangzhuoran Fin Yang}\\%
        Department of Econometrics \& Business Statistics\\%
        Monash University\\%
        Clayton VIC 3800\\%
        Australia\\%
      {Email: Fin.Yang@monash.edu}\\%
      \textit{Corresponding author}\\[0.5cm]%
      \textbf{George Athanasopoulos}\\%
        Department of Econometrics \& Business Statistics\\%
        Monash University\\%
        Clayton VIC 3800\\%
        Australia\\%
      \\[0.5cm]%
      \textbf{Rob~J Hyndman}\\%
        Department of Econometrics \& Business Statistics\\%
        Monash University\\%
        Clayton VIC 3800\\%
        Australia\\%
      \\[0.5cm]%
      \textbf{Anastasios Panagiotelis}\\%
        Discipline of Business Analytics\\%
        University of Sydney\\%
        Camperdown NSW 2050\\%
        Australia\\%
      \\[0.5cm]%
   }%
\renewenvironment{abstract}{\begin{minipage}{\textwidth}\parskip=1.4ex\noindent
\hrule\vspace{0.1cm}\par{\sffamily\textbf{\abstractname}}\newline\setstretch{1.5}}
  {\end{minipage}}
\begin{document}
\maketitle

\begin{abstract}
A novel forecast linear augmented projection (FLAP) method is
introduced, which reduces the forecast error variance of any unbiased
multivariate forecast without introducing bias. The method first
constructs new component series which are linear combinations of the
original series. Forecasts are then generated for both the original and
component series. Finally, the full vector of forecasts is projected
onto a linear subspace where the constraints implied by the combination
weights hold. It is proven that the trace of the forecast error variance
is non-increasing with the number of components, and mild conditions are
established for which it is strictly decreasing. It is also shown that
the proposed method achieves maximum forecast error variance reduction
among linear projection methods. The theoretical results are validated
through simulations and two empirical applications based on Australian
tourism and FRED-MD data. Notably, using FLAP with Principal Component
Analysis (PCA) to construct the new series leads to substantial forecast
error variance reduction.
\end{abstract}

    \vspace{0.25cm}\par\hrule\vspace{0.5cm}\par
  
\setstretch{1.5}
\section{Introduction}\label{sec-introduction}

We will show that any multivariate forecasting method can be improved
through a post-processing framework involving linear combinations of the
original series. We call this FLAP: Forecast Linear Augmented
Projection. This has immediate implications for multivariate forecasting
in all disciplines including macroeconomics \autocite{macroreview} and
finance \autocite{financereview}.

Our FLAP post-processing framework: (i) augments the data by
constructing new series that are linear combinations of the original
series; (ii) forecasts both the original and new series; and (iii)
recovers a new set of forecasts for the original series via projections.
We prove that the method reduces the forecast error variance of the
original series in a way that is agnostic both with respect to the
weights of the linear combinations used at step (i) and with respect to
the model used to generate forecasts at step (ii). The model is inspired
by the forecast reconciliation literature \autocite{AthEtAl2023a}
whereby forecasts are adjusted to cohere with known linear constraints.
In contrast to that literature, the FLAP method focuses on multivariate
forecasting where such constraints are not present. Indeed, the method
need not only be applied to forecasting problems, but multivariate
predictions in general.

It may appear puzzling that forecast accuracy can be improved, not by
introducing any new information, but by simply taking linear
combinations of existing time series. To give an intuition into how this
puzzle can be resolved, we consider a toy example of two series \(y_1\)
and \(y_2\) that are of concern to the forecaster, and two linear
combinations or \emph{components} of the series \(c_1=0.5z_1+0.5z_2\)
and \(c_2=0.5z_1-0.5z_2\). Denote by \(\hat{y}_1\), \(\hat{y}_2\),
\(\hat{c}_1\), \(\hat{c}_2\) any forecasts of these original series and
components, which we collectively refer to as \emph{base forecasts}. The
base forecasts may be generated by univariate methods, multivariate
methods, or even based on expert judgement. When considering \(y_1\),
there is both a \emph{direct} forecast \(\hat{y}_1\) and \emph{indirect}
forecast \(\hat{c}_1+\hat{c}_2\), similarly for \(y_2\) the direct
forecast is \(\hat{y}_2\) and the indirect forecast
\(\hat{c}_1-\hat{c}_2\).\footnote{This argument, as well as the
  terminology direct and indirect forecasts, is inspired by
  \textcite{HolEtAl2021} who discuss this in the forecast reconciliation
  setting.} With the exception of some pathological cases, the direct
and indirect forecasts for the same variable will not, in general, be
equal. Therefore forecast accuracy can be improved by combining direct
and indirect forecasts, something implicitly achieved by the proposed
FLAP method. The puzzle is thus resolved; while no new information is
created at the data augmentation step, there is new information embedded
into forecasts of the augmented series, which can be leveraged via model
combination (see \textcite{WanEtAl2023} for a review of forecast
combination). Something obscured by the simple toy example is the way
our FLAP method differs from the usual forecast combination methods, in
particular our combinations are potentially non-convex since they are
obtained via projections, in a way that we now elaborate upon.

More formally and more generally, the FLAP method considers a vector of
original series \(\mathbf{y}\in \mathbb{R}^m\) and a vector of
components \(\mathbf{c}\in \mathbb{R}^p\). While
\(\left(\mathbf{y}',\mathbf{c}'\right)'\) is a \(p+m\)-vector, the
construction of components as linear combinations of the original series
implies that \(\left(\mathbf{y}',\mathbf{c}'\right)'\) lies on a linear
subspace of at most dimension \(m\). The corresponding vector of
forecasts \(\left(\hat{\mathbf{y}}',\hat{\mathbf{c}}'\right)'\) will, in
general, have support on \(\mathbb{R}^{m+p}\). FLAP projects
\(\left(\hat{\mathbf{y}}',\hat{\mathbf{c}}'\right)'\) onto the
\(m\)-dimensional linear subspace on which
\(\left(\mathbf{y}',\mathbf{c}'\right)'\) has support. The setup of this
problem bears similarities to the well known problem of forecast
reconciliation where \textcite{PanEtAl2021} provide similar geometric
intuition, while \textcite{WicEtAl2019}, \textcite{AthEtAl2017}, and
\textcite{DiGir2023} have all shown that reconciliation can reduce
forecast error variance theoretically and empirically. However, we note
that these papers establish that reconciliation improves forecast
accuracy for the hierarchy \emph{as a whole}. In the general
multivariate setting that we consider, this would imply improvements in
forecast accuracy for \(\mathbf{y}\) and \(\mathbf{c}\) taken together.
This poses a problem if improvements in forecast accuracy for
\(\mathbf{c}\) could be offset by a deterioration in forecast accuracy
for \(\mathbf{y}\), since the former are not of interest in and of
themselves. A key insight we make in this paper, is that reductions in
forecast error variance accrue even for a subset of variables in the
hierarchy. It is this contribution that allows us to propose a method
that goes beyond the case where time series adhere to linear
constraints, and that instead applies to the more general setting.

While the theoretical results apply for any linear combinations of the
original series, in practice we propose to augment the data with
principal components (PCs). When doing so, the FLAP method bears a
resemblance to Dynamic Factor Models (DFMs), specifically those common
in macroeconomic forecasting
\autocite{StoWat2002a,StoWat2002,StoWat2012}, their extensions in the
machine learning literature \autocite{DeEtAl2019}, as well as the factor
augmented VAR \autocite{BerEtAl2005}. The factor models assume that the
multivariate time series possesses common components and the dynamics of
the observed series are governed by the dynamics of these unobserved
factors, often assumed to follow some parametric model. In contrast,
FLAP is a post-forecasting step; indeed, forecasts can even be made
using a DFM and then further improved by implementing the FLAP method,
something we demonstrate in Section~\ref{sec-simulation} and
Section~\ref{sec-empirical-applications}.

In the sense that forecast accuracy can be improved without any new
information, FLAP has parallels with bootstrap aggregation or
``bagging'' \autocite{Bre1996,BerEtAl2016}. Bagging can reduce
prediction variance without increasing bias \autocite{HasEtAl2003}, by
mitigating model uncertainty \autocite{PetEtAl2018}, and does so without
introducing any new data, but rather resampled versions of the existing
data. Our FLAP method also reduces forecast error variance without
introducing new data, but using linear combinations of the existing
data, rather than bootstrapping. The FLAP method improves forecast
accuracy through forecast combination and data augmentation. Other
methods that also do this include the theta method
\autocite{AssNik2000}, temporal aggregation
\autocite{KouEtAl2014,AthEtAl2017}, forecasting with sub-seasonal series
\autocite[FOSS,][]{LiEtAl2022c} and forecast combination with multiple
starting points \autocite{DisPet2015}; a review of all these methods can
be found in \textcite{PetSpi2021} who refer to them as using ``the
wisdom of data''. Our FLAP method is distinct in that it aims to exploit
information in the data with a focus on linear combinations of
multivariate series.

The remainder of the paper is structured as follows. In
Section~\ref{sec-method}, we propose the FLAP method, and highlight its
theoretical properties and associated estimation methods. In
Section~\ref{sec-simulation}, we present a simulation example
demonstrating its performance and discuss the implications for sources
of uncertainty. Section~\ref{sec-empirical-applications} examines the
performance of FLAP in two empirical applications: forecasting
Australian domestic tourism and forecasting macroeconomic variables in
the FRED-MD data set. Section~\ref{sec-conclusion} concludes with some
thoughts on future research directions. The methods introduced in this
paper are implemented in the \texttt{flap} package for R
\autocite{flap}. This paper is fully reproducible with code and
documentation provided at
\url{https://github.com/FinYang/paper-forecast-projection}.

\section{Forecast Linear Augmented Projection (FLAP)}\label{sec-method}

\subsection{Method and theoretical
properties}\label{method-and-theoretical-properties}

In the following, all vectors and matrices are denoted in bold font. We
use \(\bm{I}_n\) to denote the \(n\times n\) identity matrix, and
\(\bm{O}_{n\times k}\) to denote the \(n\times k\) zero matrix.

Let \(\bm{y}_t\in\mathbf{R}^m\) be a vector of \(m\) observed time
series we are interested in forecasting. The FLAP method involves three
steps:

\begin{enumerate}
\def\labelenumi{\arabic{enumi}.}
\tightlist
\item
  \emph{Form components}. Form
  \(\bm{c}_t = \bm{\Phi}\bm{y}_t\in\mathbf{R}^p\), a vector of \(p\)
  linear combinations of \(\bm{y}_t\) at time \(t\), where
  \(\bm{\Phi}\in\mathbf{R}^{p\times m}\). We call \(\bm{c}_t\) the
  components of \(\bm{y}_t\) and the component weights \(\bm{\Phi}\) are
  known in the sense that they are chosen by the user of FLAP. Let
  \(\bm{z}_{t} = \big[\bm{y}_t', \bm{c}'_{t}\big]'\) be the
  concatenation of series \(\bm{y}_t\) and components \(\bm{c}_{t}\). We
  note that \(\bm{z}_{t}\) will be constrained in the sense that
  \(\bm{C}\bm{z}_t= \bm{c}_{t} - \bm{\Phi}\bm{y}_t = \bm{0}\) for any
  \(t\) where \(\bm{C} = \big[- \bm{\Phi} ~~~ \bm{I}_{p}\big]\) is
  referred to as the constraint matrix.
\item
  \emph{Generate forecasts}. Denote as \(\hat{\bm{z}}_{t+h}\) the
  \(h\)-step-ahead base forecast of \(\bm{z}_{t}\). The method used to
  generate forecasts is again selected by the user. This can be
  univariate or multivariate. In the setting where \(\bm{z}_t\) are not
  time series but cross sectional data, any prediction method can be
  used. In general, the constraints that hold for \(\bm{z}_t\) will not
  hold for \(\hat{\bm{z}}_{t+h}\), i.e
  \(\bm{C}\hat{\bm{z}}_{t+h}\neq \bm{0}\)
\item
  \emph{Project the base forecasts}. Let \(\tilde{\bm{z}}_{t+h}\) be a
  set of projected forecasts such that,
  \begin{equation}\phantomsection\label{eq-z_tilde}{
  \tilde{\bm{z}}_{t+h} = \bm{M} \hat{\bm{z}}_{t+h}
  }\end{equation} with projection matrix
  \begin{equation}\phantomsection\label{eq-M}{
  \bm{M} = \bm{I}_{m+p} - \bm{W}_h\bm{C}'(\bm{C}\bm{W}_h\bm{C}')^{-1}\bm{C},
  }\end{equation} where
  \(\Var(\bm{z}_{t+h}-\hat{\bm{z}}_{t+h}) = \bm{W}_h\) is the forecast
  error covariance matrix. For the proofs of this section, we will
  assume that \(\bm{W}_h\) is known, in practice a plug-in estimate can
  be used that will be discussed in Section~\ref{sec-estW}.
\end{enumerate}

In practice we are interested in forecasts of \(\bm{y}_t\) and not the
full vector \(\bm{z}_t\). We now introduce some notation to handle this
issue. Define the selection matrix
\(\bm{J}_{n,k} = \big[\bm{I}_n ~~~ \bm{O}_{n\times k}\big]\), so that
\(\bm{J}_{n,k}\bm{A}\) selects the first \(n\) rows of a matrix
\(\bm{A}\). Let \(\hat{\bm{y}}_{t+h}\) and \(\tilde{\bm{y}}_{t+h}\)
denote the first \(m\) elements of \(\hat{\bm{z}}_{t+h}\) and
\(\tilde{\bm{z}}_{t+h}\), comprising the base and projected forecasts of
\(\bm{y}_t\) respectively. Similarly, let \(\hat{\bm{c}}_{t+h}\) and
\(\tilde{\bm{c}}_{t+h}\) denote the last \(p\) elements of
\(\hat{\bm{z}}_{t+h}\) and \(\tilde{\bm{z}}_{t+h}\), comprising the base
and projected forecasts of \(\bm{c}_t\) respectively. Then the projected
forecast of \(\bm{y}_t\) can be found by
\begin{equation}\phantomsection\label{eq-lcmap}{
\tilde{\bm{y}}_{t+h} = \bm{J}\tilde{\bm{z}}_{t+h} = \bm{J}\bm{M} \hat{\bm{z}}_{t+h},
}\end{equation} where \(\bm{J} = \bm{J}_{m,p}\).

We now present some theoretical results regarding the FLAP method, with
proofs provided in the Appendix. Theorem~\ref{thm-psdvar} establishes
that the forecasts produced by the FLAP method,
\(\tilde{\bm{y}}_{t+h}\), dominate the base forecasts,
\(\hat{\bm{y}}_{t+h}\), in the sense that the difference between their
forecast error variances is always positive semi-definite.
Theorem~\ref{thm-monotone} establishes that the trace of the covariance
of the forecast errors of \(\tilde{\bm{y}}_{t+h}\) is non-increasing
with the number of components \(p\). The conditions needed to make the
trace strictly decreasing are discussed in
Theorem~\ref{thm-pos-condition}. In Theorem~\ref{thm-minvar} we prove
that the projection in Equation~\ref{eq-M} achieves the minimum forecast
error variance amongst the class of all projections. Finally, while the
theoretical results imply that components could in principle continue to
be added to improve forecasts, in practice, larger values of \(p\) will
make estimation of the plug-in covariance matrix \(\bm{W}_h\)
unreliable. We explore this issue in a simulation setting and
empirically in Section~\ref{sec-simulation} and
Section~\ref{sec-empirical-applications}, respectively.

\subsection{Positive Semi-Definiteness of Error Variance
Reduction}\label{positive-semi-definiteness-of-error-variance-reduction}

We first provide some intermediate results.

\begin{lemma}[]\protect\hypertarget{lem-projectionM}{}\label{lem-projectionM}

Matrix \(\bm{M}\) is a projection onto the space where the constraint
\(\bm{C}\bm{z}_t=\bm{0}\) is satisfied.

\end{lemma}

Proof in Appendix, page \pageref{proof1}.

\begin{toappendix}

\subsubsection{\texorpdfstring{Proof of
Lemma~\ref{lem-projectionM}}{Proof of Lemma~}}\label{proof-of-lem-projectionm}

\label{proof1} We have \[
\begin{aligned}
\bm{M}\bm{M}
& = \bm{I}_{m+p} - 2 \bm{W}_h\bm{C}'(\bm{C}\bm{W}_h\bm{C}')^{-1}\bm{C} \\
& \mbox{}\hspace*{1cm} + \bm{W}_h\bm{C}'(\bm{C}\bm{W}_h\bm{C}')^{-1}\bm{C}\bm{W}_h\bm{C}'(\bm{C}\bm{W}_h\bm{C}')^{-1}\bm{C}\\
& =\bm{I}_{m+p} - \bm{W}_h\bm{C}'(\bm{C}\bm{W}_h\bm{C}')^{-1}\bm{C} \\
& = \bm{M},
\end{aligned}
\] so \(\bm{M}\) is a projection matrix. For any \(\bm{z}\) such that
\(\bm{M}\bm{z} = \bm{y}\) for some \(\bm{y}\), we have \[
\bm{C}\bm{y} = \bm{C}\bm{M}\bm{z} = \bm{C}\bm{z} - \bm{C}\bm{W}_h\bm{C}'(\bm{C}\bm{W}_h\bm{C}')^{-1}\bm{C}\bm{z} = \bm{0}.
\] Thus, \(\bm{M}\) projects any vector onto the space where the
constraint \(\bm{C}\bm{y}=0\) is satisfied.

\end{toappendix}

Based on the properties of projections, we have the following
corollaries.

\begin{corollary}[]\protect\hypertarget{cor-corM}{}\label{cor-corM}

~

\begin{enumerate}
\def\labelenumi{\arabic{enumi}.}
\tightlist
\item
  The projected forecast \(\tilde{\bm{z}}_{t+h}\) satisfies the
  constraint \(\bm{C}\tilde{\bm{z}}_{t+h}=\bm{0}\).
\item
  For \(\bm{z}_{t+h}\) that already satisfies the constraint, the
  projection does not change its value, i.e.,
  \(\bm{M}\bm{z}_{t+h} = \bm{z}_{t+h}\) \autocite[Lemma
  2.4]{Rao1974-JRSSSBSM}.
\item
  If the base forecasts are unbiased such that
  \(\E(\hat{\bm{z}}_{t+h}|\mathcal{I}_t) = \E(\bm{z}_{t+h}|\mathcal{I}_t)\),
  then the projected forecasts are also unbiased, i.e.,
  \(\E(\tilde{\bm{z}}_{t+h}|\mathcal{I}_t) = \E(\bm{z}_{t+h}|\mathcal{I}_t)\).
\end{enumerate}

\end{corollary}

Proof in Appendix, page \pageref{proof2}.

\begin{toappendix}

\subsubsection{\texorpdfstring{Proof of
Corollary~\ref{cor-corM}}{Proof of Corollary~}}\label{proof-of-cor-corm}

\label{proof2} Items 1 and 2 are trivial application of
Lemma~\ref{lem-projectionM}. To prove 3, we have \[
\E(\tilde{\bm{z}}_{t+h}|\mathcal{I}_t) =
\E(\bm{M}\hat{\bm{z}}_{t+h}|\mathcal{I}_t) = \bm{M}\E(\hat{\bm{z}}_{t+h}|\mathcal{I}_t) = \bm{M}\E(\bm{z}_{t+h}|\mathcal{I}_t) = \E(\bm{M}\bm{z}_{t+h}|\mathcal{I}_t) = \E(\bm{z}_{t+h}|\mathcal{I}_t).
\]

\end{toappendix}

The assumption of unbiasedness of the base forecasts is not unreasonable
in practice, and where it does not hold, a bias correction method can be
applied. Note this is not a requirement on model specification. We do
not assume the model producing the base forecast is correctly specified
like in the DFM literature \autocite[e.g.,][]{StoWat2002}. In fact, the
power of FLAP manifests when the models are misspecified, as discussed
in Section~\ref{sec-simulation}.

\begin{lemma}[]\protect\hypertarget{lem-var}{}\label{lem-var}

The forecast error covariance matrix of the component-augmented
projected \(h\)-step-ahead forecasts \(\tilde{\bm{z}}_{t+h}\) is \[
\Var(\bm{z}_{t+h} - \tilde{\bm{z}}_{t+h}) = \bm{M}\bm{W}_h\bm{M}' = \bm{M}\bm{W}_h,
\] and the forecast error covariance matrix of the projected
\(h\)-step-ahead forecasts \(\tilde{\bm{y}}_{t+h}\) is \[
\Var(\bm{y}_{t+h} - \tilde{\bm{y}}_{t+h}) = \bm{J}\bm{M}\bm{W}_h\bm{J}'.
\]

\end{lemma}

Proof in Appendix, page \pageref{proof3}.

\begin{toappendix}

\subsubsection{\texorpdfstring{Proof of
Lemma~\ref{lem-var}}{Proof of Lemma~}}\label{proof-of-lem-var}

\label{proof3} \[
\Var(\tilde{\bm{z}}_{t+h} - \bm{z}_{t+h})
= \Var(\bm{M}\hat{\bm{z}}_{t+h} - \bm{M}\bm{z}_{t+h}) \\
= \bm{M}\Var(\hat{\bm{z}}_{t+h} - \bm{z}_{t+h})\bm{M}' \\
= \bm{M}\bm{W}_h\bm{M}'.
\] If we simplify it further, we have \[
\begin{aligned}
\bm{M}\bm{W}_h\bm{M}'
&= (\bm{I} - \bm{W}_h\bm{C}'(\bm{C}\bm{W}_h\bm{C}')^{-1}\bm{C})\bm{W}_h(\bm{I} - \bm{W}_h\bm{C}'(\bm{C}\bm{W}_h\bm{C}')^{-1}\bm{C})'\\
&= \bm{W}_h - \bm{W}_h\bm{C}'(\bm{C}\bm{W}_h\bm{C}')^{-1}\bm{C}\bm{W}_h - \bm{W}_h\bm{C}'(\bm{C}\bm{W}_h\bm{C}')^{-1}\bm{C}\bm{W}_h \\
&\mbox{}\hspace*{1cm} + \bm{W}_h\bm{C}'(\bm{C}\bm{W}_h\bm{C}')^{-1}\bm{C}\bm{W}_h\bm{C}'(\bm{C}\bm{W}_h\bm{C}')^{-1}\bm{C}\bm{W}_h \\
&= \bm{W}_h - \bm{W}_h\bm{C}'(\bm{C}\bm{W}_h\bm{C}')^{-1}\bm{C}\bm{W}_h \\
&= \bm{M}\bm{W}_h.
\end{aligned}
\] To get \(\Var(\tilde{\bm{y}}_{t+h} - \bm{y}_{t+h})\), we just need to
recognise that it is the first \(m\times m\) leading principal submatrix
of \(\Var(\tilde{\bm{z}}_{t+h} - \bm{z}_{t+h})\).

\end{toappendix}

Lemma~\ref{lem-var} is a well-known result in the forecast
reconciliation literature \autocite[e.g.,][]{DiGir2023}.

\begin{theorem}[Positive Semi-Definiteness of Error Variance
Reduction]\protect\hypertarget{thm-psdvar}{}\label{thm-psdvar}

The difference between the forecast error variances of the base and
projected component-augmented forecasts, \[
\begin{aligned}
\Var(\bm{z}_{t+h}-\hat{\bm{z}}_{t+h}) -\Var(\bm{z}_{t+h}-\tilde{\bm{z}}_{t+h} ) &=\bm{W}_h-\bm{M}\bm{W}_h\\
&= \bm{W}_h - (\bm{I} - \bm{W}_h\bm{C}'(\bm{C}\bm{W}_h\bm{C}')^{-1}\bm{C})\bm{W}_h\\
&=\bm{W}_h\bm{C}'(\bm{C}\bm{W}_h\bm{C}')^{-1}\bm{C}\bm{W}_h,
\end{aligned}
\] is positive semi-definite. The difference between the forecast error
variances of the base and projected forecasts of the original series, \[
\Var(\bm{y}_{t+h}-\hat{\bm{y}}_{t+h}) -\Var(\bm{y}_{t+h}-\tilde{\bm{y}}_{t+h}) = \bm{J}\bm{W}_h\bm{C}'(\bm{C}\bm{W}_h\bm{C}')^{-1}\bm{C}\bm{W}_h\bm{J}',
\] is therefore also positive semi-definite.

\end{theorem}

Proof in Appendix, page \pageref{proof4}.

\begin{toappendix}

\subsubsection{\texorpdfstring{Proof of
Theorem~\ref{thm-psdvar}}{Proof of Theorem~}}\label{proof-of-thm-psdvar}

\label{proof4} Trivially,
\(\bm{W}_h\bm{C}'(\bm{C}\bm{W}_h\bm{C}')^{-1}\bm{C}\bm{W}_h\) and
\(\bm{J}\bm{W}_h\bm{C}'(\bm{C}\bm{W}_h\bm{C}')^{-1}\bm{C}\bm{W}_h\bm{J}'\)
are positive semi-definite. Note that
\(\Var(\hat{\bm{y}}_{t+h} - \bm{z}_{t+h}) -\Var(\tilde{\bm{y}}_{t+h} - \bm{z}_{t+h})\)
is the leading principal submatrix of
\(\bm{W}_h\bm{C}'(\bm{C}\bm{W}_h\bm{C}')^{-1}\bm{C}\bm{W}_h\), and the
leading principal submatrix of a positive semi-definite matrix is
positive semi-definite.

\end{toappendix}

Theorem~\ref{thm-psdvar} is why FLAP works. Note when
\(\bm{J}\bm{W}_h\bm{C}'(\bm{C}\bm{W}_h\bm{C}')^{-1}\bm{C}\bm{W}_h\bm{J}'\)
is semi-definite, each of its diagonal elements, the reduction in
forecast error variance for each series, is non-nagative. It implies
that the forecast error variance can be reduced by simply forecasting
the components (the artificially constructed linear combinations of the
original data), and mapping the forecasts using matrix \(\bm{M}\). For
the improvement to be zero, the trace must be zero. This implies that
\(\bm{J}\bm{W}_h\bm{C}'(\bm{C}\bm{W}_h\bm{C}')^{-1}\bm{C}\bm{W}_h\bm{J}' = \bm{O}_{m\times m}\),
as this is a positive semi-definite matrix, something rarely observed in
practice. See Theorem~\ref{thm-pos-condition} for more discussion.

The following example illustrates the mechanism of the reduction in the
forecast error variance.

\begin{example}[]\protect\hypertarget{exm-identityW}{}\label{exm-identityW}

Suppose \(\bm{z}_t\) comprises \(m\) original series \(\bm{y}_t\) and
\(p\) components \(\bm{c}_t\). Let \(\hat{\bm{z}}_{t+h}\) and
\(\tilde{\bm{z}}_{t+h}\) be \(h\)-step-ahead base and projected
forecasts of \(\bm{z}_t\). Assume that their corresponding forecast
errors are uncorrelated with unit variance such that
\(\bm{W}_h = \bm{I}_{m+p}\). Then \[
\begin{aligned}
\Var(\bm{z}_{t+h}-\hat{\bm{z}}_{t+h}) -\Var(\tilde{\bm{z}}_{t+h} - \bm{z}_{t+h})
&=
\bm{C}'(\bm{C}\bm{C}')^{-1}\bm{C}\\
& =
\begin{bmatrix} -\bm{\Phi}'\\ \bm{I}_{p} \end{bmatrix}
(\bm{\Phi}\bm{\Phi}' + \bm{I}_p)^{-1}
\begin{bmatrix} -\bm{\Phi}& \bm{I}_{p} \end{bmatrix} ,
\end{aligned}
\] where \[
\bm{C} = \begin{bmatrix} - \bm{\Phi}& \bm{I}_{p} \end{bmatrix}.
\] Let \(\bm{\Phi}\) consist of \(p\le m\) orthogonal unit vectors, for
example, those obtained from Principal Component Analysis
\autocite[PCA,][]{Jol2002}. In this case
\(\bm{\Phi}\bm{\Phi}' = \bm{I}_p\) and \[
\Var(\bm{z}_{t+h}-\hat{\bm{z}}_{t+h}) -\Var(\bm{z}_{t+h}-\tilde{\bm{z}}_{t+h})
  =  \frac{1}{2}
      \begin{bmatrix}
        \bm{\Phi}'\bm{\Phi} & -\bm{\Phi}'\\
        -\bm{\Phi} & \bm{I}_p
       \end{bmatrix}.
\] Focusing on the forecast error variance reduction of the forecasts of
the original series \(\bm{y}_t\), i.e.,
\(\tr(\Var(\bm{y}_{t+h}-\hat{\bm{y}}_{t+h}) -\Var(\bm{y}_{t+h}-\tilde{\bm{y}}_{t+h})) = \frac{1}{2}\tr(\bm{\Phi}'\bm{\Phi})\).

\begin{itemize}
\tightlist
\item
  When \(p<m\), since \(\bm{\Phi}'\bm{\Phi}\) is idempotent,
  \(\tr(\bm{\Phi}'\bm{\Phi}) = \operatorname{rank}(\bm{\Phi}'\bm{\Phi}) = p\).
  Hence, focusing on the original series the reduction in the total
  forecast error variance of the FLAP forecasts relative to the base
  forecasts, is \(p/2\).
\item
  When \(p=m\), where all principal components are used,
  \(\bm{\Phi}'\bm{\Phi} = \bm{I}_m\). This implies a total reduction in
  the error variance of \(m/2\), and that for each of the \(m\)
  individual series the error variance is halved.
\end{itemize}

If we keep increasing the number of components beyond \(m\), the result
in Theorem~\ref{thm-psdvar} still holds, although \(\bm{\Phi}\) can no
longer contain orthogonal vectors, and the example here becomes
intractable. This is an artificial example as the forecast error
variance \(\bm{W}_h\) can hardly be an identity matrix in practice. It
is likely that the forecast error of a linear combination of series will
be correlated to the forecast error of forecasting these series
directly. Nonetheless, the aim of the example is to demonstrate how the
forecast error variance can be reduced as a result of the component
forecasts bringing new information about the original series. The
forecast error variance reduction becomes larger as we increase the
number of components \(p\). This is not a coincidence but a desirable
property of FLAP, as shown in the next section.

\end{example}

\subsection{Monotonicity}\label{monotonicity}

In the results that follow, we break the base forecast error covariance
matrix into smaller blocks. \[
\bm{W}_h =
\begin{bmatrix}
\bm{W}_{y, h} & \bm{W}_{yc, h} \\
\bm{W}_{yc, h}' & \bm{W}_{c, h}
\end{bmatrix},
\] where \(\bm{W}_{y, h}\) is the forecast error covariance matrix of
\(\hat{\bm{y}}_{t+h}\), \(\bm{W}_{c, h}\) is the forecast error
covariance matrix of \(\hat{\bm{c}}_{t+h}\), and \(\bm{W}_{yc, h}\)
contains error covariances between elements of \(\hat{\bm{y}}_{t+h}\)
and \(\hat{\bm{c}}_{t+h}\).

\begin{theorem}[Monotonicity]\protect\hypertarget{thm-monotone}{}\label{thm-monotone}

The forecast error variance reductions for each series, i.e., the
diagonal elements in the matrix \[
\Var(\bm{y}_{t+h}-\hat{\bm{y}}_{t+h}) -\Var(\bm{y}_{t+h}-\tilde{\bm{y}}_{t+h}) = \bm{J}\bm{W}_h\bm{C}'(\bm{C}\bm{W}_h\bm{C}')^{-1}\bm{C}\bm{W}_h\bm{J}'
\] are non-decreasing as \(p\) increases. In particular, the sum of
forecast error variance reductions
\begin{equation}\phantomsection\label{eq-trdvar}{
\tr(\Var(\bm{y}_{t+h}-\hat{\bm{y}}_{t+h}) -\Var(\bm{y}_{t+h}-\tilde{\bm{y}}_{t+h})) = \tr(\bm{J}\bm{W}_h\bm{C}'(\bm{C}\bm{W}_h\bm{C}')^{-1}\bm{C}\bm{W}_h\bm{J}')
}\end{equation} is non-decreasing as \(p\) increases.

\end{theorem}

Proof in Appendix, page \pageref{proof5}.

\begin{toappendix}

\subsubsection{\texorpdfstring{Proof of
Theorem~\ref{thm-monotone}}{Proof of Theorem~}}\label{proof-of-thm-monotone}

\label{proof5} Suppose now that we want to include \(q\) more components
\(\bm{c}_{t}^*=\bm{\Phi}^*\bm{y}_t\) in the projection. We define
\(\bm{z}_{t}^* = \begin{bmatrix}\bm{z}_{t}\\ \bm{c}_{t}^*\end{bmatrix}\),
the constraint matrix \begin{equation}\phantomsection\label{eq-Cstar}{
\bm{C}^* =
\begin{bmatrix}
\multicolumn{2}{c}{\bm{C}} & \underset{p\times q}{\bm{0}} \\
-\underset{q\times m}{\bm{\Phi}^*} & \underset{q\times p}{\bm{0}} & \bm{I}_{q}
\end{bmatrix}
=
\begin{bmatrix}
-\underset{p \times m}{\bm{\Phi}}& \bm{I}_{p} &\underset{p\times q}{\bm{0}}\\
-\underset{q\times m}{\bm{\Phi}^*} & \underset{q\times p}{\bm{0}} & \bm{I}_{q},
\end{bmatrix}
=
\begin{bmatrix}
\overline{\bm{C}} \\ \underline{\bm{C}}
\end{bmatrix}
}\end{equation} where \(\overline{\bm{C}}\) contains the first \(p\)
rows of \(\bm{C}^*\) and \(\underline{\bm{C}}\) contains the remaining
\(q\) rows of \(\bm{C}^*\), the forecast error variance matrix \[
\Var(\hat{\bm{z}}_{t+h}^* - \bm{z}_{t+h}^*) = \bm{W}_h^* = \begin{bmatrix}\bm{W}_h & \bm{W}_{yc, h}^*\\ \bm{W}_{cy, h}^* & \bm{W}_{c,h}^*\end{bmatrix}.
\] where \(\hat{\bm{z}}_{t+h}^*\) is the \(h\)-step-ahead base forecasts
of \(\bm{z}_{t}^*\): \[
\hat{\bm{z}}_{t+h}^* = \begin{bmatrix} \hat{\bm{z}}_{t+h} \\ \hat{\bm{c}}_{t+h}^* \end{bmatrix},
\] and the corresponding \[
\bm{M}^* = \bm{I} - \bm{W}_h^*\bm{C}^{*\prime}(\bm{C}^*\bm{W}_h^*\bm{C}^{*\prime})^{-1}\bm{C}^*.
\]

Proving Theorem~\ref{thm-monotone} requires proving the following two
items.

\begin{enumerate}
\def\labelenumi{\arabic{enumi}.}
\tightlist
\item
  Including additional components in the mapping without including
  corresponding component constraints is equivalent to not including
  these additional components at all.
\item
  For a fixed set of components to be included in the mapping, adding
  constraints will reduce forecast error variance.
\end{enumerate}

We start by proving the first statement. Consider the case where we
include the additional series \(\bm{c}_{t}^*\) without using the
additional constraint \(\bm{\Phi}^*\). Defining \(\bm{M}^{+}\) only with
\(\overline{\bm{C}}\): \begin{equation}\phantomsection\label{eq-Mplus}{
\bm{M}^{+} = \bm{I}_{m+p+q} - \bm{W}_h^*\overline{\bm{C}}'(\overline{\bm{C}}\bm{W}_h^*\overline{\bm{C}}')^{-1}\overline{\bm{C}},
}\end{equation} we have
\(\tilde{\bm{z}}_{t+h}^+ = \bm{M}^{+}\hat{\bm{z}}_{t+h}^*\). Further, we
obtain \[
\bm{W}_h^*\overline{\bm{C}}'
 = \begin{bmatrix}
      \bm{W}_h & \bm{W}_{yc, h}^*\\ \bm{W}_{cy, h}^* & \bm{W}_{c, h}^*
    \end{bmatrix}
    \begin{bmatrix}
      \bm{C}'\\ \underset{q\times p}{\bm{0}} \\
    \end{bmatrix} \\
 = \begin{bmatrix}
      \bm{W}_h\bm{C}' \\
      \bm{W}_{cy, h}^*\bm{C}'
    \end{bmatrix}
\] and \[
  \overline{\bm{C}}\bm{W}_h^*\overline{\bm{C}}'
 = \begin{bmatrix}
      \bm{C} & \underset{p\times q}{\bm{0}} \\
    \end{bmatrix}
    \begin{bmatrix}
      \bm{W}_h\bm{C}' \\
      \bm{W}_{cy, h}^*\bm{C}'
    \end{bmatrix} \\
 =\bm{C}\bm{W}_h\bm{C}',
\] which gives \[
\begin{aligned}
\bm{M}^{+}
& = \bm{I}_{m+p+q} -
    \begin{bmatrix}
      \bm{W}_h\bm{C}' \\
      \bm{W}_{cy, h}^*\bm{C}'
    \end{bmatrix}
    (\bm{C}\bm{W}_h\bm{C}')^{-1}
    \begin{bmatrix}
      \bm{C} & \underset{p\times q}{\bm{0}} \\
    \end{bmatrix}\\
& = \bm{I}_{m+p+q} -
    \begin{bmatrix}
      \bm{W}_h\bm{C}'(\bm{C}\bm{W}_h\bm{C}')^{-1}\bm{C} & \bm{0} \\
      \bm{W}_{cy, h}^*\bm{C}'(\bm{C}\bm{W}_h\bm{C}')^{-1}\bm{C} & \bm{0}
    \end{bmatrix},
\end{aligned}
\] and \[
\begin{aligned}
\tilde{\bm{z}}_{t+h}^+
& = \bm{M}^{+}\hat{\bm{z}}_{t+h}^* \\
& = \left(
      \bm{I}_{m+p+q} -
      \begin{bmatrix}
        \bm{W}_h\bm{C}'(\bm{C}\bm{W}_h\bm{C}')^{-1}\bm{C} & \bm{0} \\
        \bm{W}_{cy, h}^*\bm{C}'(\bm{C}\bm{W}_h\bm{C}')^{-1}\bm{C} & \bm{0}
      \end{bmatrix}
    \right)
    \begin{bmatrix} \hat{\bm{z}}_{t+h} \\ \hat{\bm{c}}_{t+h}^* \end{bmatrix}\\
& = \begin{bmatrix}
      (\bm{I}_{n+p} - \bm{W}_h\bm{C}'(\bm{C}\bm{W}_h\bm{C}')^{-1}\bm{C})\hat{\bm{z}}_{t+h}\\
      \hat{\bm{c}}_{t+h}^* - \bm{W}_{cy, h}^*\bm{C}'(\bm{C}\bm{W}_h\bm{C}')^{-1}\bm{C}\hat{\bm{z}}_{t+h}
    \end{bmatrix} \\
& = \begin{bmatrix}
      \bm{M} \hat{\bm{z}}_{t+h}\\
      \hat{\bm{c}}_{t+h}^* - \bm{W}_{cy, h}^*\bm{C}'(\bm{C}\bm{W}_h\bm{C}')^{-1}\bm{C}\hat{\bm{z}}_{t+h}
    \end{bmatrix} \\
& = \begin{bmatrix}
      \tilde{\bm{z}}_{t+h}\\
      \hat{\bm{c}}_{t+h}^* - \bm{W}_{cy, h}^*\bm{C}'(\bm{C}\bm{W}_h\bm{C}')^{-1}\bm{C}\hat{\bm{z}}_{t+h}
    \end{bmatrix}.
\end{aligned}
\] If we only consider the forecast performance relevant to
\(\bm{y}_{t+h}\), and define
\(\bm{J}^* = \bm{J}_{m,p+q}= \begin{bmatrix}\bm{I}_{m} & \bm{0}_{m\times(p+q)} \end{bmatrix}\),
we have \[
\tilde{\bm{y}}_{t+h}^+ = \bm{J}^*\tilde{\bm{z}}_{t+h}^+ = \bm{J}\tilde{\bm{z}}_{t+h} = \tilde{\bm{y}}_{t+h}.
\] This means adding additional components without imposing the
corresponding constraints will yield the same projected forecasts as if
these additional components are not added, which implies that the
forecast error variance stays the same:
\begin{equation}\phantomsection\label{eq-acnc}{
\Var(\tilde{\bm{y}}_{t+h}^+ - \bm{y}_{t+h}) = \Var(\tilde{\bm{y}}_{t+h} - \bm{y}_{t+h}) = \bm{J}\bm{M}\bm{W}_h\bm{J}'.
}\end{equation} This finishes the proof of the first statement. Now we
move on to proving the second statement. We have the forecast error
variance matrices \[
\begin{aligned}
\Var(\tilde{\bm{z}}_{t+h}^+ - \bm{z}_{t+h}^*) & = \bm{M}^{+}\bm{W}_h^*
    = (\bm{I}_{m+p+q} -\bm{W}_h^*\overline{\bm{C}}'(\overline{\bm{C}}\bm{W}_h^*\overline{\bm{C}}')^{-1}\overline{\bm{C}})\bm{W}_h^* \\
\text{and}\qquad\qquad
\Var(\tilde{\bm{z}}_{t+h}^* - \bm{z}_{t+h}^*) & = \bm{M}^*\bm{W}_h^*
    = (\bm{I}_{m+p+q} - \bm{W}_h^*\bm{C}^{*\prime}(\bm{C}^*\bm{W}_h^*\bm{C}^{*\prime})^{-1}\bm{C}^*)\bm{W}_h^*.
\end{aligned}
\] Taking the difference, we have \[
\begin{aligned}
\Var(\tilde{\bm{z}}_{t+h}^+ - \bm{z}_{t+h}^*) - \Var(\tilde{\bm{z}}_{t+h}^* - \bm{z}_{t+h}^*)
& = (\bm{W}_h^*\bm{C}^{*\prime}(\bm{C}^*\bm{W}_h^*\bm{C}^{*\prime})^{-1}\bm{C}^* -
      \bm{W}_h^*\overline{\bm{C}}'(\overline{\bm{C}}\bm{W}_h^*\overline{\bm{C}}')^{-1}\overline{\bm{C}})\bm{W}_h^* \\
& = \bm{W}_h^*(\bm{C}^{*\prime}(\bm{C}^*\bm{W}_h^*\bm{C}^{*\prime})^{-1}\bm{C}^* -
      \overline{\bm{C}}'(\overline{\bm{C}}\bm{W}_h^*\overline{\bm{C}}')^{-1}\overline{\bm{C}})\bm{W}_h^*.
\end{aligned}
\] Using block matrix inversion, we have \[
\begin{aligned}
\bm{C}^{*\prime}(\bm{C}^*\bm{W}_h^*\bm{C}^{*\prime})^{-1}\bm{C}^*
& = \begin{bmatrix}
      \overline{\bm{C}}' & \underline{\bm{C}}'
    \end{bmatrix}
    \begin{bmatrix}
      \overline{\bm{C}}\bm{W}_h^*\overline{\bm{C}}' & \overline{\bm{C}}\bm{W}_h^*\underline{\bm{C}}' \\
      \underline{\bm{C}}\bm{W}_h^*\overline{\bm{C}}' & \underline{\bm{C}}\bm{W}_h^*\underline{\bm{C}}'
    \end{bmatrix}^{-1}
    \begin{bmatrix}
      \overline{\bm{C}}\\  \underline{\bm{C}}
    \end{bmatrix} \\
& = \begin{bmatrix}
      \overline{\bm{C}}' & \underline{\bm{C}}'
    \end{bmatrix}
    \begin{bmatrix}
      a & b \\
      c & d
    \end{bmatrix}
    \begin{bmatrix}
      \overline{\bm{C}}\\  \underline{\bm{C}}
    \end{bmatrix} \\
& =   \overline{\bm{C}}'a \overline{\bm{C}}
    + \overline{\bm{C}}'b \underline{\bm{C}}
    + \underline{\bm{C}}'c\overline{\bm{C}}
    + \underline{\bm{C}}'d\underline{\bm{C}},
\end{aligned}
\] where \[
\begin{aligned}
a &= (\overline{\bm{C}}\bm{W}_h^*\overline{\bm{C}}')^{-1} +
     (\overline{\bm{C}}\bm{W}_h^*\overline{\bm{C}}')^{-1}\overline{\bm{C}}\bm{W}_h^*\underline{\bm{C}}' \\
  & \mbox{}\hspace*{1cm}
    (\underline{\bm{C}}\bm{W}_h^*\underline{\bm{C}}' - \underline{\bm{C}}\bm{W}_h^*\overline{\bm{C}}'
    (\overline{\bm{C}}\bm{W}_h^*\overline{\bm{C}}')^{-1} \overline{\bm{C}}\bm{W}_h^*\underline{\bm{C}}')^{-1}
    \underline{\bm{C}}\bm{W}_h^*\overline{\bm{C}}'
    (\overline{\bm{C}}\bm{W}_h^*\overline{\bm{C}}')^{-1} \\
  & = (\overline{\bm{C}}\bm{W}_h^*\overline{\bm{C}}')^{-1} +
      (\overline{\bm{C}}\bm{W}_h^*\overline{\bm{C}}')^{-1}\overline{\bm{C}}\bm{W}_h^*\underline{\bm{C}}'(\underline{\bm{C}}\bm{M}^{+}\bm{W}_h^*\underline{\bm{C}}')^{-1}
      \underline{\bm{C}}\bm{W}_h^*\overline{\bm{C}}'
      (\overline{\bm{C}}\bm{W}_h^*\overline{\bm{C}}')^{-1},\\
b & = - (\overline{\bm{C}}\bm{W}_h^*\overline{\bm{C}}')^{-1}\overline{\bm{C}}\bm{W}_h^*\underline{\bm{C}}'
    (\underline{\bm{C}}\bm{W}_h^*\underline{\bm{C}}' -
     \underline{\bm{C}}\bm{W}_h^*\overline{\bm{C}}' (\overline{\bm{C}}\bm{W}_h^*\overline{\bm{C}}')^{-1}
     \overline{\bm{C}}\bm{W}_h^*\underline{\bm{C}}')^{-1}\\
  & = - (\overline{\bm{C}}\bm{W}_h^*\overline{\bm{C}}')^{-1}\overline{\bm{C}}\bm{W}_h^*\underline{\bm{C}}'
    (\underline{\bm{C}}\bm{M}^{+}\bm{W}_h^*\underline{\bm{C}}')^{-1},\\
c & = - (\underline{\bm{C}}\bm{M}^{+}\bm{W}_h^*\underline{\bm{C}}')^{-1}
      \underline{\bm{C}}\bm{W}_h^*\overline{\bm{C}}'
      (\overline{\bm{C}}\bm{W}_h^*\overline{\bm{C}}')^{-1},\\
d & = (\underline{\bm{C}}\bm{M}^{+}\bm{W}_h^*\underline{\bm{C}}')^{-1}.
\end{aligned}
\] Thus, \begin{align*}
\bm{C}^{*\prime}(\bm{C}^*\bm{W}_h^*\bm{C}^{*\prime})^{-1}\bm{C}^*
& =
\overline{\bm{C}}'
(\overline{\bm{C}}\bm{W}_h^*\overline{\bm{C}}')^{-1}
\overline{\bm{C}}\\
& \mbox{}\hspace*{1cm} +
\overline{\bm{C}}'
(\overline{\bm{C}}\bm{W}_h^*\overline{\bm{C}}')^{-1}
\overline{\bm{C}}\bm{W}_h^*\underline{\bm{C}}'
(\underline{\bm{C}}\bm{M}^{+}\bm{W}_h^*\underline{\bm{C}}')^{-1}
\underline{\bm{C}}\bm{W}_h^*\overline{\bm{C}}'
(\overline{\bm{C}}\bm{W}_h^*\overline{\bm{C}}')^{-1}
\overline{\bm{C}}\\
& \mbox{}\hspace*{1cm} -
\overline{\bm{C}}
(\overline{\bm{C}}\bm{W}_h^*\overline{\bm{C}}')^{-1}
\overline{\bm{C}}\bm{W}_h^*\underline{\bm{C}}'
(\underline{\bm{C}}\bm{M}^{+}\bm{W}_h^*\underline{\bm{C}}')^{-1}
\underline{\bm{C}}\\
& \mbox{}\hspace*{1cm} -
\underline{\bm{C}}'
(\underline{\bm{C}}\bm{M}^{+}\bm{W}_h^*\underline{\bm{C}}')^{-1}
\underline{\bm{C}}'\bm{W}_h^*\overline{\bm{C}}'
(\overline{\bm{C}}\bm{W}_h^*\overline{\bm{C}}')^{-1}
\overline{\bm{C}}\\
& \mbox{}\hspace*{1cm} +
\underline{\bm{C}}'
(\underline{\bm{C}}\bm{M}^{+}\bm{W}_h^*\underline{\bm{C}}')^{-1}
\underline{\bm{C}}\\
&=
\overline{\bm{C}}'
(\overline{\bm{C}}\bm{W}_h^*\overline{\bm{C}}')^{-1}
\overline{\bm{C}}\\
& \mbox{}\hspace*{1cm} -
\overline{\bm{C}}
(\overline{\bm{C}}\bm{W}_h^*\overline{\bm{C}}')^{-1}
\overline{\bm{C}}\bm{W}_h^*\underline{\bm{C}}'
(\underline{\bm{C}}\bm{M}^{+}\bm{W}_h^*\underline{\bm{C}}')^{-1}
\underline{\bm{C}}\bm{M}^{+}\\
& \mbox{}\hspace*{1cm} +
\underline{\bm{C}}'
(\underline{\bm{C}}\bm{M}^{+}\bm{W}_h^*\underline{\bm{C}}')^{-1}
\underline{\bm{C}}\bm{M}^{+}\\
& =
\overline{\bm{C}}'
(\overline{\bm{C}}\bm{W}_h^*\overline{\bm{C}}')^{-1}
\overline{\bm{C}}
+
\bm{M}^{+\prime}
\underline{\bm{C}}'
(\underline{\bm{C}}\bm{M}^{+}\bm{W}_h^*\underline{\bm{C}}')^{-1}
\underline{\bm{C}}\bm{M}^{+}.
\end{align*} Therefore, \[
\begin{aligned}
\Var(\tilde{\bm{z}}_{t+h}^+ - \bm{z}_{t+h}^*) - \Var(\tilde{\bm{z}}_{t+h}^* - \bm{z}_{t+h}^*)
\hspace*{-4cm} \\
& =
\bm{W}_h^*(
\overline{\bm{C}}'
(\overline{\bm{C}}\bm{W}_h^*\overline{\bm{C}}')^{-1}
\overline{\bm{C}}
+
\bm{M}^{+\prime}
\underline{\bm{C}}'
(\underline{\bm{C}}\bm{M}^{+}\bm{W}_h^*\underline{\bm{C}}')^{-1}
\underline{\bm{C}}\bm{M}^{+}
-\overline{\bm{C}}'(\overline{\bm{C}}\bm{W}_h^*\overline{\bm{C}}')^{-1}\overline{\bm{C}})\bm{W}_h^*\\
& =
\bm{W}_h^*(
\bm{M}^{+\prime}
\underline{\bm{C}}'
(\underline{\bm{C}}\bm{M}^{+}\bm{W}_h^*\underline{\bm{C}}')^{-1}
\underline{\bm{C}}\bm{M}^{+}
)\bm{W}_h^*
\end{aligned}
\] is positive semi-definite. This concludes the proof of the second
statement. Combining the results above, we have
\begin{equation}\phantomsection\label{eq-addcom}{
\begin{aligned}
\Var(\tilde{\bm{y}}_{t+h} - \bm{y}_{t+h}) -
\Var(\tilde{\bm{y}}_{t+h}^* - \bm{y}_{t+h})
& =
\Var(\tilde{\bm{y}}_{t+h}^+ - \bm{y}_{t+h}) -
\Var(\tilde{\bm{y}}_{t+h}^* - \bm{y}_{t+h}) \\
&=
\bm{J}^*\Var(\tilde{\bm{z}}_{t+h}^+ - \bm{z}_{t+h}^*)\bm{J}^{*\prime} -
\bm{J}^*\Var(\tilde{\bm{z}}_{t+h}^* - \bm{z}_{t+h}^*)\bm{J}^{*\prime} \\
&=
\bm{J}^*
\bm{W}_h^*
\bm{M}^{+\prime}
\underline{\bm{C}}'
(\underline{\bm{C}}\bm{M}^{+}\bm{W}_h^*\underline{\bm{C}}')^{-1}
\underline{\bm{C}}\bm{M}^{+}
\bm{W}_h^*
\bm{J}^{*\prime}
\end{aligned}
}\end{equation} being positive semi-definite. Finally, we have
\begin{multline*}
(\Var(\hat{\bm{y}}_{t+h} - \bm{y}_{t+h}) -\Var(\tilde{\bm{y}}_{t+h}^* - \bm{y}_{t+h}))
-
(\Var(\hat{\bm{y}}_{t+h} - \bm{y}_{t+h}) -\Var(\tilde{\bm{y}}_{t+h} - \bm{y}_{t+h})) \\
=
    \Var(\tilde{\bm{y}}_{t+h} - \bm{y}_{t+h}) -
    \Var(\tilde{\bm{y}}_{t+h}^* - \bm{y}_{t+h})
\end{multline*} being a positive semi-definite matrix where the diagonal
terms are non-negative, whose trace, therefore, is non-negative. This
means using a larger number of components in the mapping achieves lower
or equal forecast error variances. In other words, the reduction in
forecast error variance of each series is non-decreasing as more
components are added, giving Theorem~\ref{thm-monotone}.

\end{toappendix}

Theorem~\ref{thm-monotone} is the key result that demonstrates the
usefulness of FLAP. It means that we can keep increasing the number of
components to reduce forecast error variance, even when the number of
components exceeds the number of original series. It requires \(\bm{C}\)
to be \(\big[-\bm{\Phi} ~~~ \bm{I}_{p} \big]\) or
\(\big[-\bm{\Phi} ~~~ \bm{L}\big]\) where \(\bm{L}\) is a lower
triangular matrix. This implies that the components can also be
constructed from existing components, not only from the original series.
This has little significance since a linear combination of components of
the original series, is just a linear combination of the original
series. This of course assumes that the forecast error covariances are
known, something we explore in Section~\ref{sec-estW} and
Section~\ref{sec-simulation}.

Extending the proof of Theorem~\ref{thm-monotone}, we can state the
condition required for the reduction sum of forecast error variances to
be strictly positive. Denote \(\bm{\phi}_i\) as the row vector
containing the weights associated with the \(i\)th component, so that
with \(p\) components, the weights matrix is
\(\bm{\Phi} = \big[\bm{\phi}_1' ~~~ \bm{\phi}_2' ~~~ \dots ~~~ \bm{\phi}_p'\big]'\).
Let \(\bm{W}^{(i-1)}_{\tilde{\bm{y}},h}\) denote the error covariance
matrix of the projected forecasts of the original series based on the
first \(i-1\) components, \(\bm{w}_{c_{1}\hat{\bm{y}},h}\) denote a
vector of covariances of the first component and the base forecasts of
the original series, and \(\bm{w}_{c_i\tilde{y},h}^{(i-1)}\) denote a
vector of covariances of the projected \(i\)th component and the
projected forecasts of the original series, based on the first \(i-1\)
components.

\begin{theorem}[Positive Forecast Error Variance Reduction
Condition]\protect\hypertarget{thm-pos-condition}{}\label{thm-pos-condition}

For the first component to achieve a guaranteed reduction of forecast
error variance (for the matrix in Theorem~\ref{thm-psdvar} to have
strictly positive trace),
\begin{equation}\phantomsection\label{eq-pcon}{
\bm{\phi}_1\bm{W}_{y,h}\neq\bm{w}_{c_1y,h}.
}\end{equation} For the \(i\)th component to have a positive reduction
on forecast error variance of the original series,
\begin{equation}\phantomsection\label{eq-pconi}{
\bm{\phi}_i{\bm{W}}^{(i-1)}_{\tilde{y},h}\neq\bm{w}_{c_i\tilde{y},h}^{(i-1)}.
}\end{equation}

\end{theorem}

Proof in Appendix, page \pageref{proof6}.

\begin{toappendix}

\subsubsection{\texorpdfstring{Proof of
Theorem~\ref{thm-pos-condition}}{Proof of Theorem~}}\label{proof-of-thm-pos-condition}

\label{proof6} Denote
\(\bm{\psi}_i=\begin{bmatrix}-\bm{\phi}_i & \bm{0}_{1\times(i-1)} & 1\end{bmatrix}\)
and \(\bm{W}_{h}^{(i)}\) to be the base forecast error variance of the
original series and the first \(i\) components. Starting with the first
component, Equation~\ref{eq-trdvar} becomes
\begin{equation}\phantomsection\label{eq-trdvar1}{
\tr(\bm{J}_{m,1}\bm{W}_h^{(1)}\bm{\psi}_1'(\bm{\psi}_1\bm{W}_h^{(1)}\bm{\psi}_1')^{-1}\bm{\psi}_1\bm{W}_h^{(1)}\bm{J}_{m,1}') =
(\bm{\psi}_1\bm{W}_h^{(1)}\bm{\psi}_1')^{-1}\bm{\psi}_1\bm{W}_h^{(1)}\bm{J}_{m,1}'\bm{J}_{m,1}\bm{W}_h^{(1)}\bm{\psi}_1',
}\end{equation} \[
\begin{aligned}
\text{where}\qquad\qquad
\bm{\psi}_1\bm{W}_h^{(1)}\bm{J}_{m,1}' =&
\begin{bmatrix}
-\bm{\phi}_1 & 1
\end{bmatrix}
\begin{bmatrix}
\bm{W}_{z,h} & \bm{w}'_{c_1z,h} \\
\bm{w}_{c_1z,h} & \bm{W}_{c_1,h}
\end{bmatrix}
\begin{bmatrix}
\bm{I}_m\\ 0
\end{bmatrix} \\
=& -\bm{\phi}_1\bm{W}_{z,h} + \bm{w}_{c_1z,h}.
\end{aligned}
\] Equation~\ref{eq-trdvar1} is obviously non-negative. For it to be
larger than \(0\), we need
\(\bm{\psi}_1\bm{W}_h^{(1)}\bm{J}_{m,1}' \neq 0\), which gives
\(\bm{\phi}_1\bm{W}_{z,h} \neq \bm{w}_{c_1z,h}\).

When it comes to adding the \(i\)th component on top of the first
\(i-1\) components, we define \[
\overline{\bm{C}}_i =
\begin{bmatrix}
\bm{\psi}_1 & \bm{0}_{1\times i} \\
\bm{\psi}_2 & \bm{0}_{1\times (i-1)} \\
\vdots & \vdots \\
\bm{\psi}_i & 0 \\
\end{bmatrix}
\] and \[
\bm{M}^+_i = \bm{I}_{m+i} - \bm{W}_h^{(i)} \overline{\bm{C}}_{i-1}'
(\overline{\bm{C}}_{i-1}\bm{W}_h^{(i)}\overline{\bm{C}}_{i-1}')^{-1}
\overline{\bm{C}}_{i-1}
\] analogously to Equation~\ref{eq-Cstar} and Equation~\ref{eq-Mplus}.
Following Equation~\ref{eq-addcom}, the additional reduction of forecast
error variance when adding the \(i\)th component becomes \[
\bm{J}_{m,i}
\bm{W}_h^{(i)}
\bm{M}^{+\prime}_i
\bm{\psi}_i'
(\bm{\psi}_i\bm{M}^{+}_i\bm{W}_h^{(i)}\bm{\psi}_i')^{-1}
\bm{\psi}_i\bm{M}^{+}_i
\bm{W}_h^{(i)}
\bm{J}_{m,i}'
=
(\bm{\psi}_i\bm{M}^{+}_i\bm{W}_h^{(i)}\bm{\psi}_i')^{-1}
\bm{\psi}_i
\bm{M}^{+}_i
\bm{W}_h^{(i)}
\bm{J}_{m,i}'
\bm{J}_{m,i}
\bm{W}_h^{(i)}
\bm{M}^{+\prime}_i
\bm{\psi}_i'.
\] Similar to before, we would want
\(\bm{\psi}_i\bm{M}^{+}_i\bm{W}_h^{(i)}\bm{J}_{m,i}' \neq \bm{0}\). Note
that \(\bm{\psi}_i\) concerns the first \(m\) rows and the last row of
\(\bm{M}^{+}_i\bm{W}_h^{(i)}\), and \(\bm{J}_{m,i}'\) concerns the first
\(m\) columns. Combined with the implication from Equation~\ref{eq-acnc}
that the \(m\times m\) leading principal submatrix in equation
\(\bm{J}_{m,i}\bm{M}^{+}_i\bm{W}_h^{(i)}\bm{J}_{m,i}'=\bm{J}_{m,i-1}\bm{M}_{i-1}\bm{W}_h^{(i-1)}\bm{J}_{m,i-1}'\)
is the same, we suppress the straightforward yet tiresome details, and
obtain \[
\bm{\phi}_i\bm{W}_{\tilde{z},h}^{(i-1)} \neq \big[\bm{0}_{1\times m+i-1} ~~~ 1 \big]\bm{M}^{+}_i\bm{W}_h^{(i)}\bm{J}_{m,i}',
\] where
\(\bm{W}_{\tilde{z},h}^{(i-1)}=\bm{J}_{m,i-1}\bm{M}_{i-1}\bm{W}_h^{(i-1)}\bm{J}_{m,i-1}'\)
is the projected forecast error variance of the original series using
the first \(i-1\) components, and the right hand side of the inequality
is simply a one-row matrix consisting of the first \(m\) elements in the
last row of \(\bm{M}^{+}_i\bm{W}_h^{(i)}\), which can be denoted as
\(\bm{w}_{\tilde{c}_i\tilde{z},h}^{(i-1)}\) and interpreted as the
covariance between the projected forecast of the original series using
the first \(i-1\) components, and the projected forecast of the \(i\)th
component using the first \(i-1\) components.

\end{toappendix}

The condition of Equation~\ref{eq-pcon} demonstrates that for a new
component to be beneficial, the information introduced by this new
component, reflected in the error covariance, cannot be a linear
combination of already existing information.

Theorem~\ref{thm-pos-condition} can potentially provide insights into
the selection of component weights and forecast models to satisfy the
conditions. Heuristically, we want the components to be introducing new
information, and this is more likely when they are uncorrelated, which
motivates out choice of principal components which are uncorrelated by
constructions. Optimal choices of weights we leave to future research,
but note that practically speaking, the conditions in
Theorem~\ref{thm-pos-condition} are either almost always satisfied if
the weights are simulated randomly on a continuous scale, or the loss
associated with the rare occasions where the conditions are not
satisfied is negligible compared to the estimation error imposed by the
limited sample size as the number of components increases, as discussed
in Section~\ref{sec-simulation} and
Section~\ref{sec-empirical-applications}.

\subsection{Optimality of the
projection}\label{optimality-of-the-projection}

Equation~\ref{eq-z_tilde} can be seen as a solution to the optimisation
problem, \[
\underset{\check{\bm{z}}_{T+h}}{\arg\min}
  (\hat{\bm{z}}_{T+h}-\check{\bm{z}}_{T+h})'\bm{W}_h^{-1}(\hat{\bm{z}}_{T+h}-\check{\bm{z}}_{T+h})
\qquad \text{s.t. } \bm{C}\check{\bm{z}}_{T+h}=0.
\] If we consider the transformed space where all the vectors are first
transformed via pre-multiplying by \(\bm{W}_h^{-1/2}\), where
\(\bm{W}_h^{-1} = (\bm{W}_h^{-1/2})'\bm{W}_h^{-1/2}\), then this
optimisation problem can be interpreted as finding the set of forecasts
that are closest to the base forecasts on the transformed space, while
satisfying the linear constraints imposed by the components.

An alternative way of characterising the same constraints is \[ 
\bm{\Phi}\check{\bm{y}}_{T+h}=\check{\bm{c}}_{T+h},
\] where \(\check{\bm{c}}_{T+h}\) is the vector of the last \(p\)
elements of \(\check{\bm{z}}_{T+h}\), corresponding to the forecast of
the components as part of the solution. This equivalence is discussed in
\textcite{WicEtAl2019}, where the authors find the solution by
minimising the sum of forecast error variance of all series (See
\textcite{AndNar2024-F} for an alternative proof). The result is
\begin{equation}\phantomsection\label{eq-mint}{
\tilde{\bm{z}}_{t+h} = \bm{S}\bm{G} \hat{\bm{z}}_{t+h},
}\end{equation} where
\(\bm{S} = \begin{bmatrix}\bm{I}_m \\\bm{\Phi}\end{bmatrix}\) contains
the constraints, so that \(\bm{z}_{t} = \bm{S}\bm{y}_{t}\),
and\pagebreak[3]\begin{equation}\phantomsection\label{eq-G}{
\bm{G} = (\bm{S}'\bm{W}_h^{-1}\bm{S})^{-1}\bm{S}'\bm{W}_h^{-1}.
}\end{equation} In Equation~\ref{eq-mint}, \(\bm{G} \hat{\bm{z}}_{t+h}\)
can be viewed as mapping all of the series to a selected few. In the
forecast reconciliation context, this is a mapping of all series to the
``bottom level''. In our multivariate forecasting context, this is a
mapping of all series including the components, to the space of the
original series. This leads to the solution
\begin{equation}\phantomsection\label{eq-stmap}{
\tilde{\bm{y}}_{t+h} = \bm{G}\hat{\bm{z}}_{t+h},
}\end{equation} as equivalent to Equation~\ref{eq-lcmap}. Recognising
that Equation~\ref{eq-mint} is equivalent to Equation~\ref{eq-z_tilde},
it is the solution that minimises the sum of forecast error variances of
the original series and all the components. We go further in
Theorem~\ref{thm-minvar}, and show that Equation~\ref{eq-mint} is also
the solution to minimise each individual forecast error variance of the
original series, and their sum. This can be viewed as a special case of
Theorem 3.3 in \textcite{PanEtAl2021} applied in a forecast projection
context, by taking
\(\bm{W}=\bm{S}(\bm{S}'\bm{S})^{-1}(\bm{S}'\bm{S})^{-1}\bm{S}'\) in
\textcite[Thm. 3.3]{PanEtAl2021}, as illustrated by
\textcite{AndNar2024-F}. The earliest work we can find that noted this
interpretation in a non-forecasting context is
\textcite[p.~85]{Lue1969}. We establish a few basic results leading to
the optimality of this solution first, also to check that
Lemma~\ref{lem-projectionM}, Corollary~\ref{cor-corM} and
Lemma~\ref{lem-var} hold under this alternative representation.

\begin{lemma}[]\protect\hypertarget{lem-projectionG}{}\label{lem-projectionG}

The matrix \(\bm{S}\bm{G}\) is a projection onto the space where the
constraint \(\bm{C}\bm{z}_t=\bm{0}\) is satisfied, provided that
\(\bm{G}\bm{S}=\bm{I}\).

\end{lemma}

Proof in Appendix, page \pageref{proof7}.

\begin{toappendix}

\subsubsection{\texorpdfstring{Proof of
Lemma~\ref{lem-projectionG}}{Proof of Lemma~}}\label{proof-of-lem-projectiong}

\label{proof7} If \(\bm{G}\bm{S}=\bm{I}\), \(\bm{S}\bm{G}\) is a
projection matrix: \(\bm{S}\bm{G}\bm{S}\bm{G} = \bm{S}\bm{G}\).

For any \(\bm{z}\) such that \(\bm{S}\bm{G}\bm{z} = \bm{y}\) for some
\(\bm{y}\), we have \(\bm{C}\bm{y} = \bm{C}\bm{S}\bm{G}\bm{z} = \bm{0}\)
because
\(\bm{C}\bm{S} = \big[-\bm{\Phi} ~~~ \bm{I} \big] \big[\bm{I} ~~~ \bm{\Phi}' \big]' = \bm{0}\).
Similarly to \(\bm{M}\), \(\bm{S}\bm{G}\) projects a vector to the same
space where \(\bm{C}\) is satisfied.

\end{toappendix}

\begin{corollary}[]\protect\hypertarget{cor-corG}{}\label{cor-corG}

Provided that \(\bm{G}\bm{S} = \bm{I}\), the following results hold.

\begin{enumerate}
\def\labelenumi{\arabic{enumi}.}
\tightlist
\item
  The projected forecast in Equation~\ref{eq-stmap} satisfies the
  constraint
  \(\bm{C}\tilde{\bm{z}}_{t+h}= \bm{C}\bm{S}\tilde{\bm{y}}_{t+h}= \bm{0}\).
\item
  For \(\bm{z}_{t+h}\) that already satisfies the constraint, the
  mapping does not change its value, i.e.,
  \(\bm{G}\bm{z}_{t+h} = \bm{y}_{t+h}\).
\item
  If the base forecasts are unbiased such that
  \(\E(\hat{\bm{z}}_{t+h}|\mathcal{I}_t) = \E(\bm{z}_{t+h}|\mathcal{I}_t)\),
  then the projected forecasts in Equation~\ref{eq-stmap} are also
  unbiased:
  \(\E(\tilde{\bm{y}}_{t+h}|\mathcal{I}_t) = \E(\bm{y}_{t+h}|\mathcal{I}_t)\).
\end{enumerate}

\end{corollary}

Proof in Appendix, page \pageref{proof8}.

\begin{toappendix}

\subsubsection{\texorpdfstring{Proof of
Corollary~\ref{cor-corG}}{Proof of Corollary~}}\label{proof-of-cor-corg}

\label{proof8} Item 1 is an direct application of
Lemma~\ref{lem-projectionG}. From Lemma~\ref{lem-projectionG} and Lemma
2.4 in \textcite{Rao1974-JRSSSBSM}, we have \[
\bm{S}\bm{G}\bm{z}_{t+h} = \bm{z}_{t+h} = \bm{S}\bm{y}_{t+h}.
\] Left multiplying by \(\bm{G}\) on both sides, we have
\(\bm{G}\bm{z}_{t+h} = \bm{y}_{t+h}\) and item 2 is proven. To prove
Item 3, we have \[
\E(\tilde{\bm{y}}_{t+h}|\mathcal{I}_t) =
\E(\bm{G}\hat{\bm{z}}_{t+h}|\mathcal{I}_t) = \bm{G}\E(\hat{\bm{z}}_{t+h}|\mathcal{I}_t) = \bm{G}\E(\bm{z}_{t+h}|\mathcal{I}_t) = \E(\bm{G}\bm{z}_{t+h}|\mathcal{I}_t) = \E(\bm{y}_{t+h}|\mathcal{I}_t).
\]

\end{toappendix}

\begin{lemma}[]\protect\hypertarget{lem-st-var}{}\label{lem-st-var}

The covariance matrix of the projected forecasts from
Equation~\ref{eq-stmap} is given by \[
\Var(\bm{y}_{t+h} - \tilde{\bm{y}}_{t+h}) = \bm{G}\bm{W}_h\bm{G}'.
\]

\end{lemma}

Proof in Appendix, page \pageref{proof9}.

\begin{toappendix}

\subsubsection{\texorpdfstring{Proof of
Lemma~\ref{lem-st-var}}{Proof of Lemma~}}\label{proof-of-lem-st-var}

\label{proof9} Let the base and projected forecast errors be given as \[
\begin{aligned}
\hat{\bm{e}}_{y, t+h} &= \bm{y}_{t+h} - \hat{\bm{y}}_{t+h},\\
\hat{\bm{e}}_{z, t+h} &= \bm{z}_{t+h} - \hat{\bm{z}}_{t+h},\\
\tilde{\bm{e}}_{y, t+h} &= \bm{y}_{t+h} - \tilde{\bm{y}}_{t+h},\\
\text{and}\qquad
\tilde{\bm{e}}_{z, t+h} &= \bm{z}_{t+h} - \tilde{\bm{z}}_{t+h}
= \bm{S}\bm{y}_{t+h} - \bm{S}\tilde{\bm{y}}_{t+h}
= \bm{S} \tilde{\bm{e}}_{y, t+h}. \\
\text{Then we have}\qquad
\tilde{\bm{e}}_{z, t+h}
&= \hat{\bm{e}}_{z, t+h} + \hat{\bm{z}}_{t+h} - \tilde{\bm{z}}_{t+h}\\
&= \hat{\bm{e}}_{z, t+h} + \hat{\bm{z}}_{t+h} - \bm{S}\bm{G}\hat{\bm{z}}_{t+h}\\
&= \hat{\bm{e}}_{z, t+h} + (\bm{I} - \bm{S}\bm{G})(\bm{z}_{t+h} - \hat{\bm{e}}_{z, t+h})\\
\text{and}\qquad
&= \bm{S}\bm{G}\hat{\bm{e}}_{z, t+h} + (\bm{I} - \bm{S}\bm{G})\bm{S}\bm{y}_{t+h}\\
\bm{S} \tilde{\bm{e}}_{y, t+h} &= \bm{S}\bm{G}\hat{\bm{e}}_{z, t+h},
\end{aligned}
\] where the last line comes from \(\bm{G}\bm{S} = \bm{I}\). Left
multiplying by \(\bm{G}\) on both sides, we have \[
\bm{G}\bm{S} \tilde{\bm{e}}_{y, t+h} = \bm{G}\bm{S}\bm{G}\hat{\bm{e}}_{z, t+h}
\qquad\text{and}\qquad
\tilde{\bm{e}}_{y, t+h} = \bm{G}\hat{\bm{e}}_{z, t+h},
\] and therefore \[
\Var(\tilde{\bm{y}}_{t+h} - \bm{y}_{t+h}) =\Var(\tilde{\bm{e}}_{y, t+h})=\Var(\bm{G}\hat{\bm{e}}_{z, t+h})= \bm{G}\Var(\hat{\bm{e}}_{z, t+h})\bm{G}'= \bm{G}\bm{W}_h\bm{G}'.
\]

\end{toappendix}

We are now ready to present the following theorem.

\begin{theorem}[Minimum Variance Unbiased Projected
Forecast]\protect\hypertarget{thm-minvar}{}\label{thm-minvar}

The solution to \begin{equation}\phantomsection\label{eq-obj}{
\underset{\bm{G}}{\arg\min}\ \tr(\bm{G}\bm{W}_h\bm{G}')
\qquad \text{s.t. } \bm{G}\bm{S} = \bm{I}
}\end{equation} is Equation~\ref{eq-G}. This problem can be effectively
split into independent subproblems such that
\(\bm{G} = \big[\bm{g}_1 ~~ \bm{g}_2 ~~ \dots ~~ \bm{g}_m\big]'\), where
\(\bm{g}_i\) is the solution to the subproblem of the \(i\)th series
\begin{equation}\phantomsection\label{eq-subobj}{
\underset{\bm{g}_i}{\arg\min}\ \bm{g}_i'\bm{W}_h\bm{g}_i
\qquad \text{s.t. } \bm{g}_i'\bm{s}_{j} = \delta_{ij}, \quad j= 1, 2, .\ldots, m,
}\end{equation} where \(\bm{s}_j\) is the \(j\)th column of \(\bm{S}\),
and \(\delta_{ij}\) is the Kronecker delta function taking value 1 if
\(i = j\) and 0 otherwise.

\end{theorem}

Proof in Appendix, page \pageref{proof10}.

\begin{toappendix}

\subsubsection{\texorpdfstring{Proof of
Theorem~\ref{thm-minvar}}{Proof of Theorem~}}\label{proof-of-thm-minvar}

\label{proof10} This can be proved in a few different ways. We adopt the
approach of \textcite{AndNar2024-F} to obtain the solution to
Equation~\ref{eq-obj}, but the procedure from \textcite[p.~85]{Lue1969}
can also be used, where the problem is divided to
Equation~\ref{eq-subobj} and reconstructed to find the solution to
Equation~\ref{eq-obj}.

There exists a Lagrange multiplier \(\bm{\Lambda}\) such that \[
L(\bm{G}) = \tr(\bm{G}\bm{W}_h\bm{G}') + \tr(\bm{\Lambda}'(\bm{I} - \bm{G}\bm{S}))
\] is stationary at an extremum \(\bm{G}\) \autocite[p.~243, Theorem
1]{Lue1969}. We find the numerator of the Gateaux differential
\autocite[p.~171]{Lue1969} \[
\lim_{\alpha \to 0} \frac{L(\bm{G} + \alpha\bm{H}) - L(\bm{G})}{\alpha}
\] to be \[
\begin{aligned}
L(\bm{G} + \alpha\bm{H}) - L(\bm{G}) =&
\tr((\bm{G} + \alpha\bm{H})\bm{W}_h(\bm{G} + \alpha\bm{H})') + \tr(\bm{\Lambda}'(\bm{I} - (\bm{G} + \alpha\bm{H})\bm{S})) - \tr(\bm{G}\bm{W}_h\bm{G}') - \tr(\bm{\Lambda}'(\bm{I} - \bm{G}\bm{S})) \\
=&
\tr(\bm{G}\bm{W}_h\bm{G}') + \tr(\alpha^2\bm{H}\bm{W}_h\bm{H}') + \tr(\alpha\bm{G}\bm{W}_h\bm{H}') + \tr(\alpha\bm{H}\bm{W}_h\bm{G}') + \tr(\bm{\Lambda}'(\bm{I} - \bm{G}\bm{S})) - \tr(\alpha\bm{\Lambda}'\bm{H}\bm{S}) \\
&- \tr(\bm{G}\bm{W}_h\bm{G}') - \tr(\bm{\Lambda}'(\bm{I} - \bm{G}\bm{S}))\\
=& \alpha^2\tr(\bm{H}\bm{W}_h\bm{H}') + \alpha\tr(\bm{G}\bm{W}_h\bm{H}') + \alpha\tr(\bm{H}\bm{W}_h\bm{G}')  -\alpha\tr(\bm{\Lambda}'\bm{H}\bm{S}).
\end{aligned}
\] Thus, the Gateaux differential becomes \[
\begin{aligned}
\lim_{\alpha \to 0} \frac{L(\bm{G} + \alpha\bm{H}) - L(\bm{G})}{\alpha} =&
\lim_{\alpha \to 0} \alpha\tr(\bm{H}\bm{W}_h\bm{H}') + \tr(\bm{G}\bm{W}_h\bm{H}') + \tr(\bm{H}\bm{W}_h\bm{G}')  -\tr(\bm{\Lambda}'\bm{H}\bm{S})\\
=& 
\tr(\bm{G}\bm{W}_h\bm{H}') + \tr(\bm{H}\bm{W}_h\bm{G}') - \tr(\bm{\Lambda}'(\bm{H}\bm{S}))\\
=&\tr(2\bm{H}\bm{W}_h\bm{G}'-\bm{\Lambda}'\bm{H}\bm{S})\\
=& \tr(\bm{H}(2\bm{W}_h\bm{G}'-\bm{S}\bm{\Lambda}')),
\end{aligned}
\] which we set to zero with a value of \(\bm{G}^*\) \[
\begin{aligned}
\tr(\bm{H}(2\bm{W}_h\bm{G}^{*\prime}-\bm{S}\bm{\Lambda}'))&=0 \\
2\bm{W}_h\bm{G}^* &= \bm{S}\bm{\Lambda}' \\
\bm{G}^{*\prime} &= \frac{1}{2}\bm{W}_h^{-1}\bm{S}\bm{\Lambda}'.
\end{aligned}
\] Multiplying \(\bm{S}'\) to the left of both sides. we have \[
\bm{S}'\bm{G}^{*\prime}=\bm{I} = \frac{1}{2}\bm{S}'\bm{W}_h^{-1}\bm{S}\bm{\Lambda}'
\qquad\text{and}\qquad
\bm{\Lambda}' = 2 (\bm{S}'\bm{W}_h^{-1}\bm{S})^{-1}
\] because \(\bm{G}^*\bm{S} = \bm{I}\). Putting it back in, we have \[
\bm{G}^{*\prime} = \bm{W}_h^{-1}\bm{S}(\bm{S}'\bm{W}_h^{-1}\bm{S})^{-1}
\qquad\text{and}\qquad
\bm{G}^* = (\bm{S}'\bm{W}_h^{-1}\bm{S})^{-1}\bm{S}'\bm{W}_h^{-1}.
\]

To see how Equation~\ref{eq-obj} can be split into separate problems,
recognise \[
\tr(\bm{G}\bm{W}_h\bm{G}') = \sum^m_{i=1} \bm{g}_i'\bm{W}_h\bm{g}_i,
\] where \(\bm{W}_h\) is a positive definite variance covariance matrix,
which makes each \(\bm{g}_i'\bm{W}_h\bm{g}_i\) positive. Additionally,
the element in the \(i\)th row and \(j\)th column of \(\bm{G}\bm{S}\) is
\(\bm{g}_i'\bm{s}_{j}\), and the element in the \(i\)th row and \(j\)th
column of an identity matrix is \(\delta_{ij}\). Therefore, the problem
in Equation~\ref{eq-obj} is \(m\) separate problems.

\end{toappendix}

In other words, the forecast projection method gives optimal projected
forecast for a given set of components, in the sense that the unbiased
forecast of each series has minimum variance.

\subsection{\texorpdfstring{Estimation of
\(\bm{W}_h\)}{Estimation of \textbackslash bm\{W\}\_h}}\label{sec-estW}

In practice, the base forecast error variance \(\bm{W}_h\) is unknown
and needs to be estimated. Denote
\(\hat{\bm{e}}_{t,h} = \bm{z}_{t} - \hat{\bm{z}}_{t|t-h}\) as the
\(h\)-step-ahead base forecast in-sample residual. The conventional
forecast error variance matrix estimator \[
\widehat{\bm{W}}_h=\frac{1}{T-h-1}\sum^T_{t=h+1}\hat{\bm{e}}_{t,h}\hat{\bm{e}}_{t,h}',
\] albeit unbiased, is not considered a good approximation to the true
forecast error variance in a finite sample when \((m+p) \approx T-h\).
It is even singular when \((m+p)>T-h\), which makes the quantities
discussed in the previous sections impossible to calculate. For this
reason, while other shrinkage estimators can be considered, we adopt the
covariance shrinkage method of \textcite{SchStr2005} and the variance
shrinkage method of \textcite{OpgStr2007}. Then, the estimated forecast
error variance matrix is guaranteed to be positive definite with no
numerical problems when computing their inverse. This estimator is
denoted as
\(\widehat{\bm{W}}_h^{shr} = (\hat{w}_{ij,h}^{shr})_{1\leq i,j\leq m+p}\)
with the element in row \(i\) and column \(j\) \[
\hat{w}_{ij,h}^{shr} = \hat{r}_{ij,h}^{shr}\sqrt{\hat{v}_{i,h}\hat{v}_{j,h}},
\] where
\(\hat{r}_{ij,h}^{shr} = (1-\hat{\lambda}_{cor})\hat{r}_{ij,h}\) and
\(\hat{v}_{i,h} = \hat{\lambda}_{var}\hat{w}_{h, median} + (1-\hat{\lambda}_{var})\hat{w}_{i, h}\),
with \(\hat{\lambda}_{cor}\) being the shrinkage intensity parameter for
the correlation \[
\hat{\lambda}_{cor} =
\min\left(1,
\frac
{\sum_{i\neq j}\widehat{\operatorname{var}}(\hat{r}_{ij,h})}
{\sum_{i\neq j}\hat{r}_{ij,h}^2}
\right),
\] and \(\hat{\lambda}_{var}\) being the shrinkage intensity parameter
for the variance \[
\hat{\lambda}_{var} =
\min\left(1,
\frac
{\sum_{i=1}^{m+p}\widehat{\operatorname{var}}(\hat{w}_{i,h})}
{\sum_{i=1}^{m+p}(\hat{w}_{i, h} - \hat{w}_{h, median})^2}
\right),
\] \(\hat{r}_{ij,h}\) the sample correlation of the \(h\)-step-ahead
forecast error between the \(i\)th and the \(j\)th series (component) in
\(\bm{z}_t\), \(\hat{w}_{i, h}\) the \(h\)-step-ahead sample base
forecast error variance associated with the \(i\)th series (the \(i\)th
diagonal element of \(\widehat{\bm{W}}_h\)), and \(\hat{w}_{h, median}\)
the median of the \(h\)-step-ahead sample forecast error variance of the
series and components (the median of the diagonal elements of
\(\widehat{\bm{W}}_h\)). The estimation of \(\widehat{\bm{W}}_h^{shr}\)
in the following sections are implemented using the package
\texttt{corpcor} \autocite{corpcor} in R \autocite{R}.

Estimating \(\widehat{\bm{W}}_h^{shr}\) for each forecast horizon \(h\)
is desirable but computationally intensive. It involves the calculation
of multi-step-ahead in-sample residuals of the forecast models, which is
especially challenging for iterative forecasts. Because of this, in
practice it is not unreasonable to assume the \(h\)-step forecast error
variance is proportional to the \(1\)-step forecast error variance by a
constant \(\eta_h\), as do \textcite{WicEtAl2019}: \[
\widehat{\bm{W}}_h^{shr} = \eta_h\widehat{\bm{W}}_1^{shr}.
\] Under this assumption, when \(\widehat{\bm{W}}_h^{shr}\) is used in
Equation~\ref{eq-M}, the proportionality constant \(\eta_h\) cancels out
regardless of the value of \(h\). We can effectively use only the
one-step forecast error variance in forecast projection, if we only need
point forecasts. We calculate \(\widehat{\bm{W}}_h^{shr}\) for each
\(h\) for the simulation example in Section~\ref{sec-simulation}, but
assume this proportionality for the application in
Section~\ref{sec-empirical-applications}.

\section{Simulation}\label{sec-simulation}

In this section, we illustrate the performance of FLAP in a simulation
setting. We generate time series of length \(T=400\) from a \(m=70\)
variable VAR(\(3\)) data generating process (DGP). The coefficients for
the VAR DGP are estimated from the first \(70\) series in the Australian
tourism data set used in Section~\ref{sec-australian-domestic-tourism}.
The innovations are simulated from a multivariate normal distribution
with an identity covariance matrix. The estimation and simulation is
performed using the \texttt{tsDyn} package \autocite{tsDyn}.

For each simulated sample we generate \(h=1\) to \(12\)-step-ahead base
forecasts from benchmark models. We implement FLAP with a varying number
of components and component construction methods. This process is
repeated \(220\) times. The improvements of FLAP forecasts over base
forecasts is assessed, and the statistical significance of these
improvements is evaluated.

\subsection{Generating base and FLAP forecasts}\label{sec-benchmarks}

We generate base forecasts from two benchmarks. The first benchmark is
the univariate ARIMA model. For each series, we generate base forecasts
from an ARIMA model using the \texttt{auto.arima()} function from the
\texttt{forecast} package \autocite{forecast} with the default settings.

The second benchmark is the dynamic factor model (DFM). Following
\textcite{StoWat2002a}, base forecasts \[
\hat{y}_{T+h} = \hat{\alpha}_h + \sum^n_{j=1}\hat{\bm{\beta}}_{hj}'\hat{\bm{F}}_{T-j+1} + \sum^s_{j=1}\hat{\gamma}_{hj}y_{T-j+1},
\] where \(\hat{\bm{F}}_t\) is the vector of \(k\) estimated factors,
and \(\hat{y}_t\) is the target series to forecast. The factors are
estimated using PCA on demeaned and scaled data. The optimal model is
selected for each series based on the Bayesian information criterion
(BIC) from models fitted using different combinations of meta-parameters
in their corresponding range: \(1 \leq k \leq 6\), \(1 \leq n \leq 3\)
and \(1 \leq s \leq 3\). We note that the DFM produces direct forecasts
in the sense that a different model is fitted for each forecast horizon
\(h\), in contrast to indirect or iterative forecasts generated by the
ARIMA models.

We use several sets of weights to construct linear components for the
FLAP method. The types of components are listed in Table~\ref{tbl-comp}.
Principal components from PCA are established using the
\texttt{prcomp()} function in the \texttt{stats} package \autocite{R}.
We generate random components using weights simulated from a standard
normal (Norm) distribution and a uniform (Unif) distribution with range
\((-1, 1)\). We normalise the weights of all randomly generated
components into unit vectors to ensure numerically stable computation,
with \(\bm{\phi}_i/\sqrt{\sum_j(\phi_{ij}^2)}\) where \(\phi_{ij}\) is
the \(j\)th value in the weight vector of the \(i\)th component. The
total number of components \(p\) is selected to be either \(m=70\) or
\(300\).

\begin{longtable}[]{@{}
  >{\raggedright\arraybackslash}p{(\columnwidth - 2\tabcolsep) * \real{0.1667}}
  >{\raggedright\arraybackslash}p{(\columnwidth - 2\tabcolsep) * \real{0.8333}}@{}}
\caption{Component construction methods for
FLAP}\label{tbl-comp}\tabularnewline
\toprule\noalign{}
\begin{minipage}[b]{\linewidth}\raggedright
Component
\end{minipage} & \begin{minipage}[b]{\linewidth}\raggedright
Description
\end{minipage} \\
\midrule\noalign{}
\endfirsthead
\toprule\noalign{}
\begin{minipage}[b]{\linewidth}\raggedright
Component
\end{minipage} & \begin{minipage}[b]{\linewidth}\raggedright
Description
\end{minipage} \\
\midrule\noalign{}
\endhead
\bottomrule\noalign{}
\endlastfoot
PCA & Principal components from PCA. \\
Norm & The weights of all components are simulated from a standard
normal distribution. \\
Unif & The weights of all components are simulated from a uniform
distribution \\
PCA+Norm & \(m\) principal components from PCA are complemented with
random components whose weights are simulated from a standard normal
distribution. \\
PCA+Unif & \(m\) principal components from PCA are complemented with
random components whose weights are simulated from a uniform
distribution. \\
Ortho & A random orthonormal matrix is generated using package
\texttt{pracma} \autocite{pracma} as the weight matrix. \\
Ortho+Norm & A random orthonormal matrix is generated using package
\texttt{pracma}, forming the weights of the first \(m\) components.
Weights of the additional components are simulated from a standard
normal distribution. \\
\end{longtable}

\subsection{Forecast evaluation}\label{sec-evaluation}

\begin{figure}

\centering{

\includegraphics{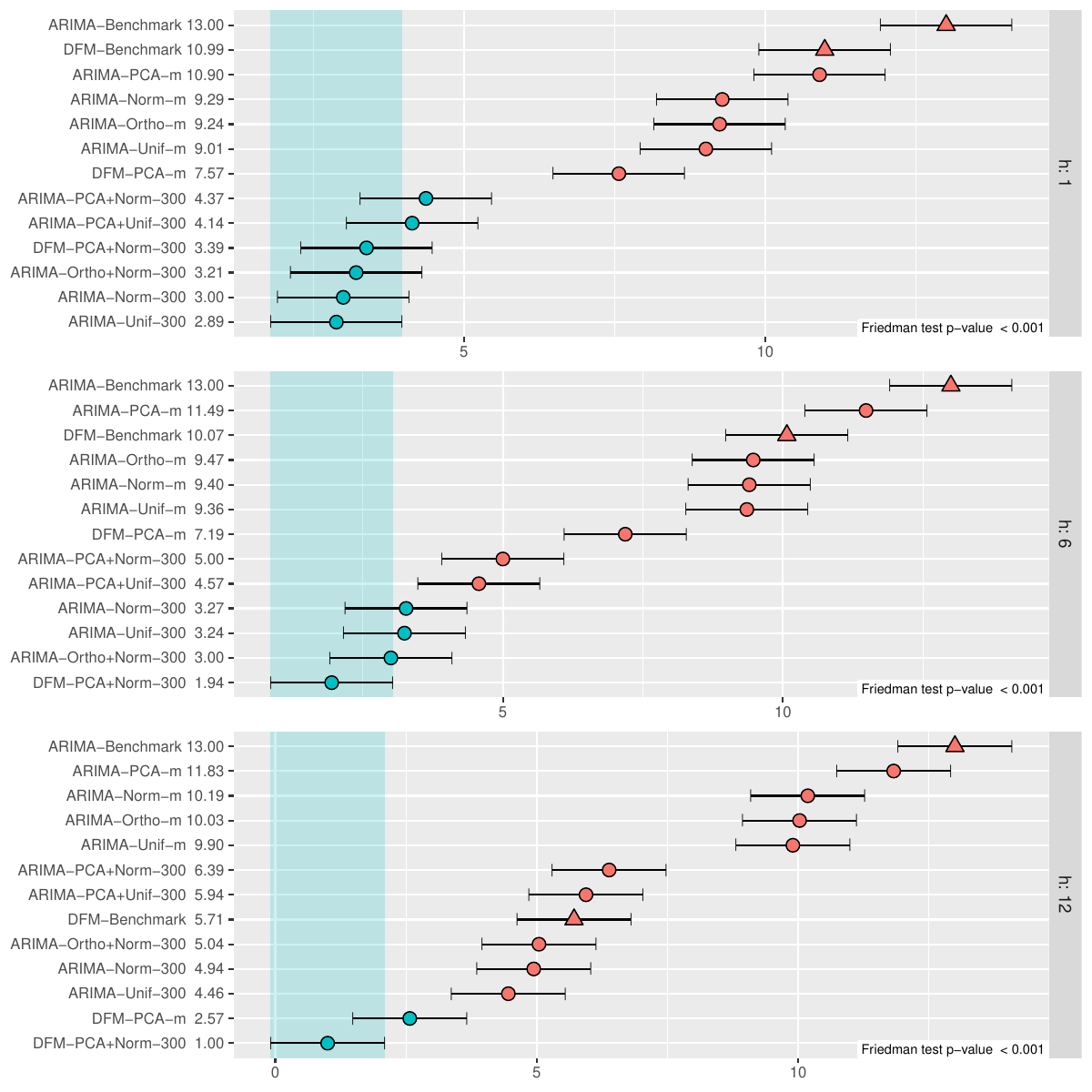}

}

\caption{\label{fig-simulation-mcb-series}Average ranks of \(1\)-,
\(6\)- and \(12\)-step-ahead MSE of different model and component
specifications in the simulation. The methods using FLAP are labelled as
``Model -- Component Weights -- Number of Components''. The two
benchmark models generating base forecasts are labelled ``Model --
Benchmark''. Their MSEs are marked with triangles. The shaded region is
the confidence interval of the best performing method. Methods outside
the shaded region are significantly worse than the best method.}

\end{figure}

We employ the Friedman test \autocite{Fri1937,Fri1939-JASA} along with
post-hoc Nemenyi tests \autocite{Nem1963,HolEtAl2013} using the
\texttt{tsutils} package \autocite{tsutils} to compare forecast
performance between different methods. The analysis involves the use of
Multiple Comparisons with the Best (MCB) plot introduced by
\textcite{KonEtAl2005} to visualise the comparison. The MSE of each
series over different samples is calculated, and the MSEs of all the
series are treated as observations in the Nemenyi test. The average
ranks are plotted in Figure~\ref{fig-simulation-mcb-series} for forecast
horizons \(1\), \(6\) and \(12\). The methods using FLAP are labelled
``Model -- Component Weights -- Number of Components''. The benchmarks
are labelled ``Model -- Benchmark''. These points are marked with
triangles. The shaded region is the confidence interval of the
best-performing method. Methods outside the shaded region are
significantly worse than the best model.

\begin{figure}

\centering{

\includegraphics{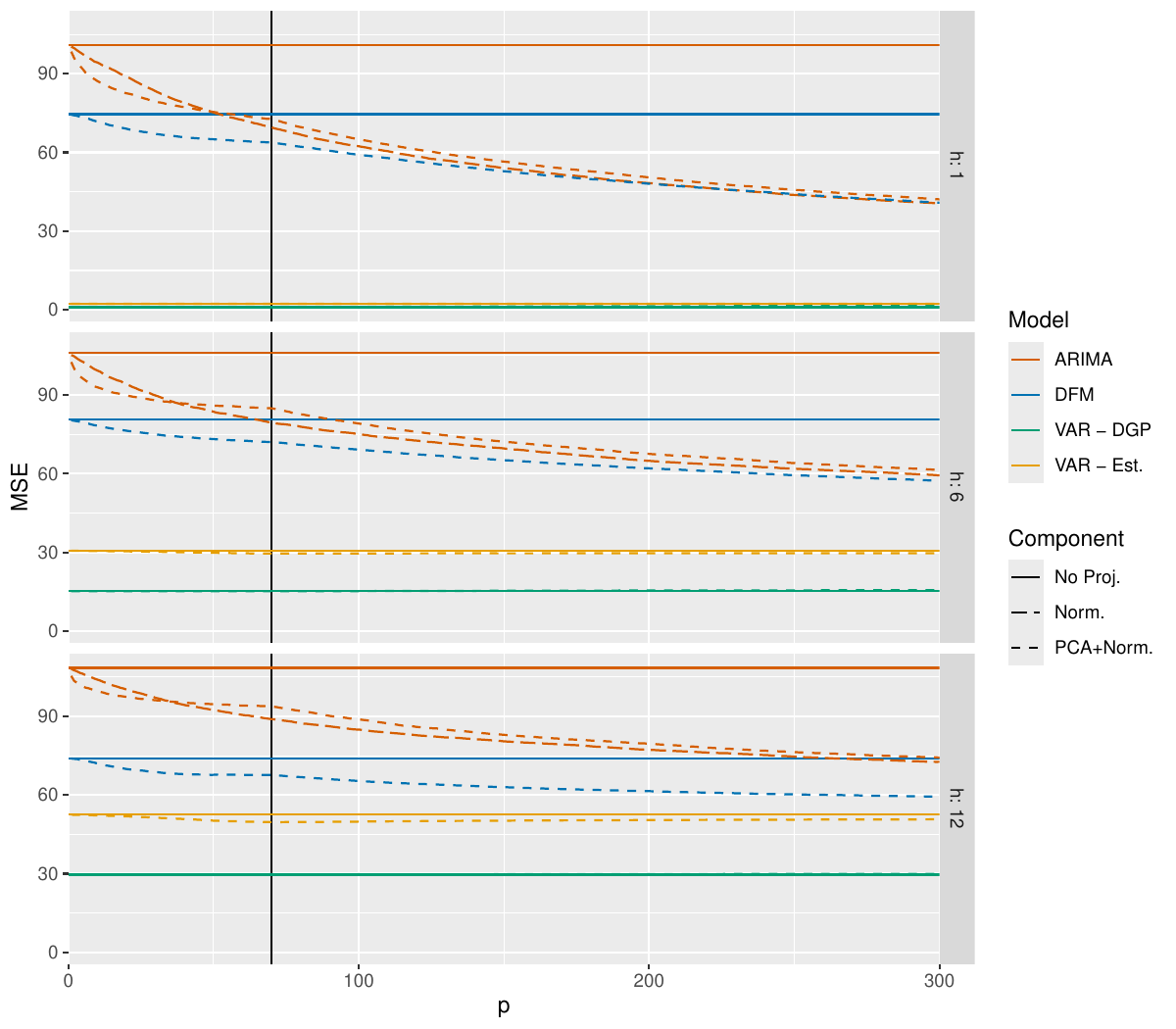}

}

\caption{\label{fig-simulation-line}Out-of-sample MSE for base and FLAP
forecasts as the number of components \(p\) increases, for forecast
horizons \(1\), \(6\) and \(12\). ``VAR -- DGP'' indicates the
performance of the true data generating VAR model. ``VAR -- Est.''
indicates the performance of the VAR model with the same structure as
the true model but with estimated parameters. The solid horizontal lines
show the MSE for the base forecasts (with no projection) while the
dashed lines show the MSEs of the FLAP forecasts. The vertical black
line indicates the location of \(p=m=70\), the number of series.}

\end{figure}

In Figure~\ref{fig-simulation-line}, we plot the out-of-sample MSE
values as a function of the number of components \(p\) included in the
FLAP method. Here we also include the performance of the DGP VAR model
(VAR -- DGP), and an estimated VAR model with the correct specification
(VAR -- Est.), as well as their corresponding FLAP generated forecasts.
We have not presented the FLAP forecasts using components generated from
a uniform distribution or random orthonormal matrices, as these are
visually identical to the methods with random weights generated from a
standard normal distribution. The vertical black line indicates when
\(p=m=70\), the number of original series.

\subsubsection{Evaluating base
forecasts}\label{evaluating-base-forecasts}

Having simulated data from a VAR DGP, we expect the DFM to capture the
correlations between series, something not possible with univariate
ARIMA models. Hence, we expect the DFM benchmark to perform better than
the ARIMA benchmark. This is indeed the case.
Figure~\ref{fig-simulation-mcb-series} shows that the base DFM forecasts
are significantly better than base ARIMA forecasts, with the exception
for \(h=1\), where the difference in rank is statistically insignificant
but only marginally.

This is also observed in Figure~\ref{fig-simulation-line}. The solid
horizontal line representing the MSE of the base DFM forecasts is always
much lower than the horizontal solid line of the base ARIMA forecasts.

\subsubsection{Evaluating FLAP
forecasts}\label{evaluating-flap-forecasts}

The most important observation from
Figure~\ref{fig-simulation-mcb-series} are the statistically significant
improvements of the FLAP forecasts over their corresponding benchmark
base forecasts. Note that we are referring here to forecasts
corresponding to the same model, i.e., FLAP ARIMA forecasts, compared to
base ARIMA forecasts and FLAP DFM forecasts, compared to base DFM
forecasts. The average ranks of the FLAP forecasts are better than the
corresponding base forecasts for all forecast horizons, and the
differences are all statistically significant. The only exception is for
ARIMA base forecasts with FLAP using only the \(p=m=70\) principal
components, where the difference in rank is statistically insignificant.

Hence, the number of components seems to be important. The
best-performing models all comprise the maximum \(p=300\) components.
For \(h=1\)-step-ahead forecasts, the differences between the methods
with \(p=300\) components are not significantly significant, regardless
of the model that generated the base forecasts or the method used to
construct the components.

Indeed, Figure~\ref{fig-simulation-line} shows that the MSEs for the
FLAP ARIMA and DFM forecasts, represented by the dashed lines, decrease
monotonically relative to the base forecasts, as the number of
components increases. This supports Theorem~\ref{thm-monotone},
demonstrating that increasing the number of components in FLAP reduces
the forecast error variance. Of course this is only feasible in this
ideal setting where each time series has \(400\) observations and we use
no more than \(300\) components in FLAP. The relatively large number of
observations coupled with the relatively simple VAR DGP, has potentially
eased the challenge of estimation. This continuous reduction in forecast
error variance is not always achievable with real data, as demonstrated
in the empirical applications in
Section~\ref{sec-empirical-applications}, especially with FRED-MD
dataset in Section~\ref{sec-fred-md}.

Worth noting from Figure~\ref{fig-simulation-line} is the performance of
the FLAP ARIMA forecasts relative to DFM forecasts as the number of
components \(p\) increases. The MSE of the FLAP ARIMA forecasts becomes
lower than the DFM base forecasts for \(p<70\) for \(h=1\), for
\(70<p<100\) for \(h=6\) and for a larger \(p\) for \(h=12\). Hence,
FLAP is able to enhance univariate ARIMA forecasts with shared
information between series by capturing and projecting these
cross-correlations with the components.

Interestingly, for \(h=1\), the MSEs of FLAP ARIMA and FLAP DFM
forecasts converge to the same value as \(p\) reaches \(300\), no matter
how the components are constructed. For the longer forecast horizons,
the MSEs again converge; they do not reach the same value for \(p=300\),
although they are closer for \(h=6\) than \(h=12\). The reason for this
is not the contribution of FLAP, but the starting point of the two sets
of base forecasts. As the forecast horizon increases from \(h=1\) to
\(6\) and \(12\) the ARIMA base forecasts MSE increases. In contrast the
MSEs for DFM base forecasts remains at a relatively constant level.
Thus, it seems that the direct forecasts from the DFM model,
particularly fitting a different model for each forecast horizon \(h\),
is advantageous for the longer forecast horizons, in this setting of
forecasting data generated from a simple VAR DGP. This is in contrast to
the iterative nature of ARIMA generated forecasts.

Note that the same forecasts of the components, generated from
univariate ARIMA models, are used for both the FLAP ARIMA and the FLAP
DFM methods. It is the increase in the number of the components, that
leads to the dominant performance of the FLAP forecasts. Hence, there is
valuable information in the series that is not captured by either the
ARIMA model or the DFM, but is captured by the components. Once again,
this emphasises the importance of the components and FLAP. In this
extreme case, a simple model after FLAP, performs as well as a more
complicated model, while the forecast model itself is not as important
as FLAP.

\subsection{Forming components for
FLAP}\label{sec-component-construction}

The construction of components is obviously a primary element in the
proposed FLAP method. However, the simulation results show that the
method of generating the components may not be as important as one may
have expected.

The most important feature in Figure~\ref{fig-simulation-mcb-series}, as
discussed in the previous section, is the number of components used in
this simulation setting. The results clearly show that the FLAP
forecasts using the maximum considered \(p=300\) components outrank the
FLAP forecasts using \(p=m=70\) components, no matter how the components
are generated. The only exception seems to be for \(h=12\) and for the
DFM base forecasts when using all \(p=m=70\) PCA components or
complementing these with randomly generated components from normal
weights. Our conjecture, as also stated above, is that the key
difference here lies in the quality of the direct DFM base forecasts for
the longer forecast horizons, rather than in the component generating
mechanism itself. Furthermore, between the methods using the maximum
considered \(p=300\), there seems to be no evidence that the component
generating mechanism makes any difference to the ranks, and especially
there seems to be no difference between the distribution used to
generate the random components.

Hence, in Figure~\ref{fig-simulation-line}, we present only the results
for principal components and randomly generated components from a
standard normal distribution. The results from using components
generated from a uniform distribution or random orthonormal matrices are
omitted, as these are visually identical to the latter.

Focusing on the ARIMA base forecasts it seems that for a relatively
small \(p\), the MSE decreases at a faster rate when principal
components are used compared to random components. However, FLAP with
random components seems to achieve a lower MSE than PCA as \(p\)
increases. The gap between the two diminishes for large \(p\). We
conjecture that the initial favourable performance of PCA stems from the
construction of the orthogonal components aimed at capturing the maximum
variance in the data. These components are then ranked from largest to
smallest variance, which likely contributes to the method's
effectiveness for smaller \(p\) compared to random components. We
comment more on PCA when analysing the results in
Section~\ref{sec-empirical-applications}.

The results show that the performance of FLAP using orthonormal weights
is almost identical to FLAP using weights simulated from a normal
distribution. This leads us to infer that it is not the orthogonality of
the principal components that leads to the strong performance of FLAP
using PCA. It may also suggest using simple random weights if one is
prepared to include a relatively large number of components in FLAP; and
using PCA when the intention is to use a small number of components.
However, as we will see in Section~\ref{sec-empirical-applications},
this is not the case with real data. In this case, PCA seems to be the
preferred approach even when the number of components is large.

\subsection{Sources of uncertainty}\label{sources-of-uncertainty}

The MSEs of the true VAR data generating process, VAR-DGP, and the
correctly specified but estimated VAR, VAR-Est, are plotted at the
bottom of each panel in Figure~\ref{fig-simulation-line}. As expected,
this confirms that they are the best performing models. As the forecast
horizon increases so does the MSE difference between these, with
estimation error for the VAR-Est accumulating as forecasts are
iteratively generated for the longer forecast horizons.

For both these models, FLAP cannot improve on the base forecasts. More
importantly, in the case of VAR-Est, FLAP does not seem to be able to
reduce the estimation error or the parameter uncertainty. On the other
hand, FLAP shows significant improvements over base forecasts generated
from misspecified models, such as the two benchmarks. This implies that
the uncertainty which FLAP can account for and reduce, is mainly due to
model misspecification. It seems that FLAP operates similarly to
bagging, as bagging also reduces variance by controlling model
uncertainty \autocite{PetEtAl2018}. The uncertainty in how the
components are constructed, unique in FLAP problems, is discussed in
Section~\ref{sec-component-construction} and awaits future research.

\section{Empirical applications}\label{sec-empirical-applications}

In this section, we apply the FLAP method in an empirical setting using
two diverse datasets. The analysis confirms many of the results and
conclusions drawn from the simulation setting, but it also identifies a
few key differences.

\subsection{Australian domestic
tourism}\label{sec-australian-domestic-tourism}

The Australian Tourism Data Set compiled from the National Visitor
Survey by Tourism Research Australia contains the total number of nights
spent by Australians away from home. We refer to these as visitor
nights. The monthly visitor nights are recorded for \(m=77\) regions
around Australia, covering the period January 1998 to December 2019. To
evaluate the forecast performance, we conduct am expanding window time
series cross-validation \autocite{fpp3}. More specifically, the first
\(T = 84\) observations are used as the first training set and the
following \(12\) months are used as the test set for evaluation. We
repeat the evaluation for the rest of the data, by expanding each
training sample by one observation at a time. This generates \(169\)
forecasts for evaluation, for each of the forecast horizons, \(h=1\) to
\(12\). Base forecasts for each series and for the components are
generated using univariate ETS models selected and fitted using the
\texttt{ets()} function in the \texttt{forecast} package
\autocite{forecast,HynKha2008}. See \textcite[Chapter 7]{fpp2} for more
details.

\begin{figure}[!b]

\centering{

\includegraphics{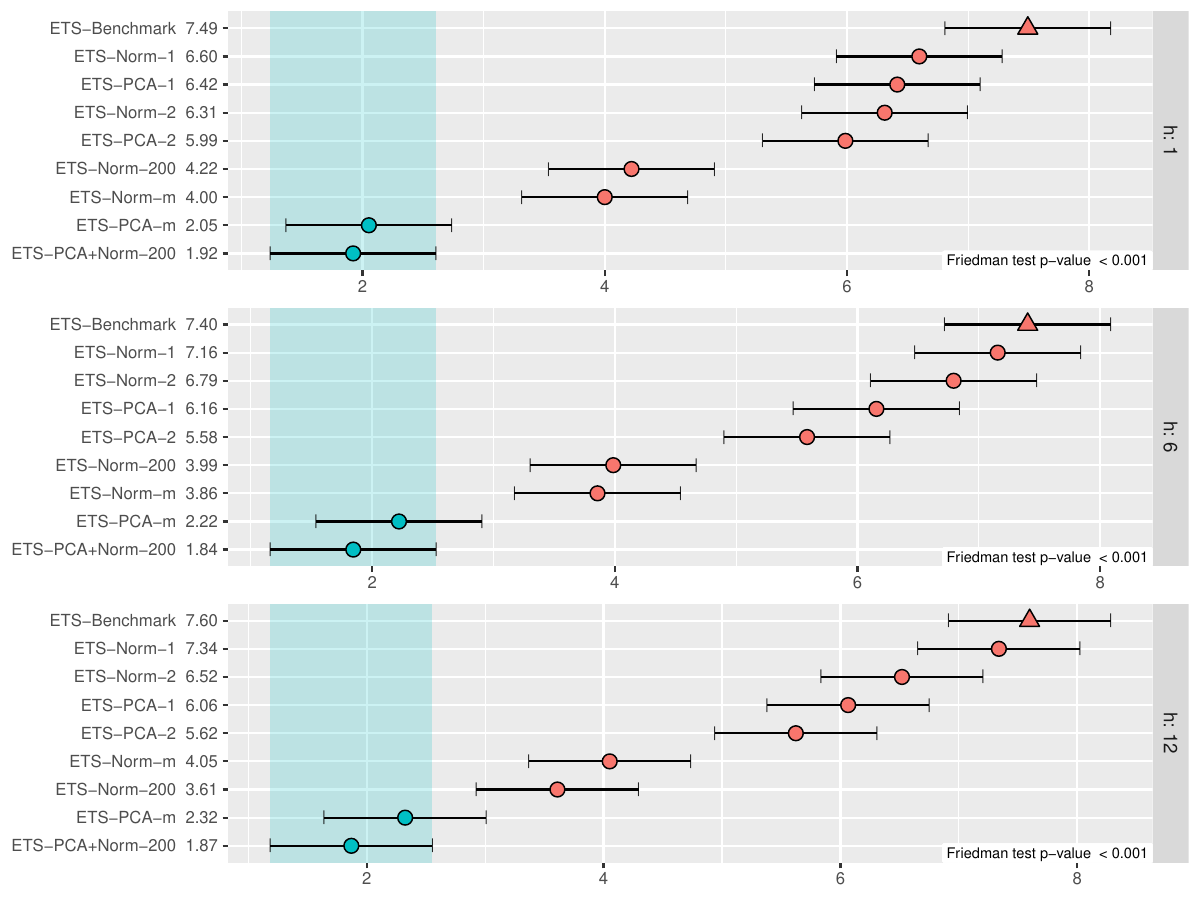}

}

\caption{\label{fig-visnights-mcb-series}Average ranks of \(1\)-, \(6\)-
and \(12\)-step-ahead cross-validation MSE of different model and
component specifications on the visitor nights data. The methods using
forecast projection are named as ``Model -- Component Weights -- Number
of Components''. The base models are named as ``Model -- Benchmark'' and
these points are marked with triangles. The shaded region is the
confidence interval of the best performing model. Methods outside the
shaded region are significantly worse than the best model.}

\end{figure}%

\begin{figure}

\centering{

\includegraphics{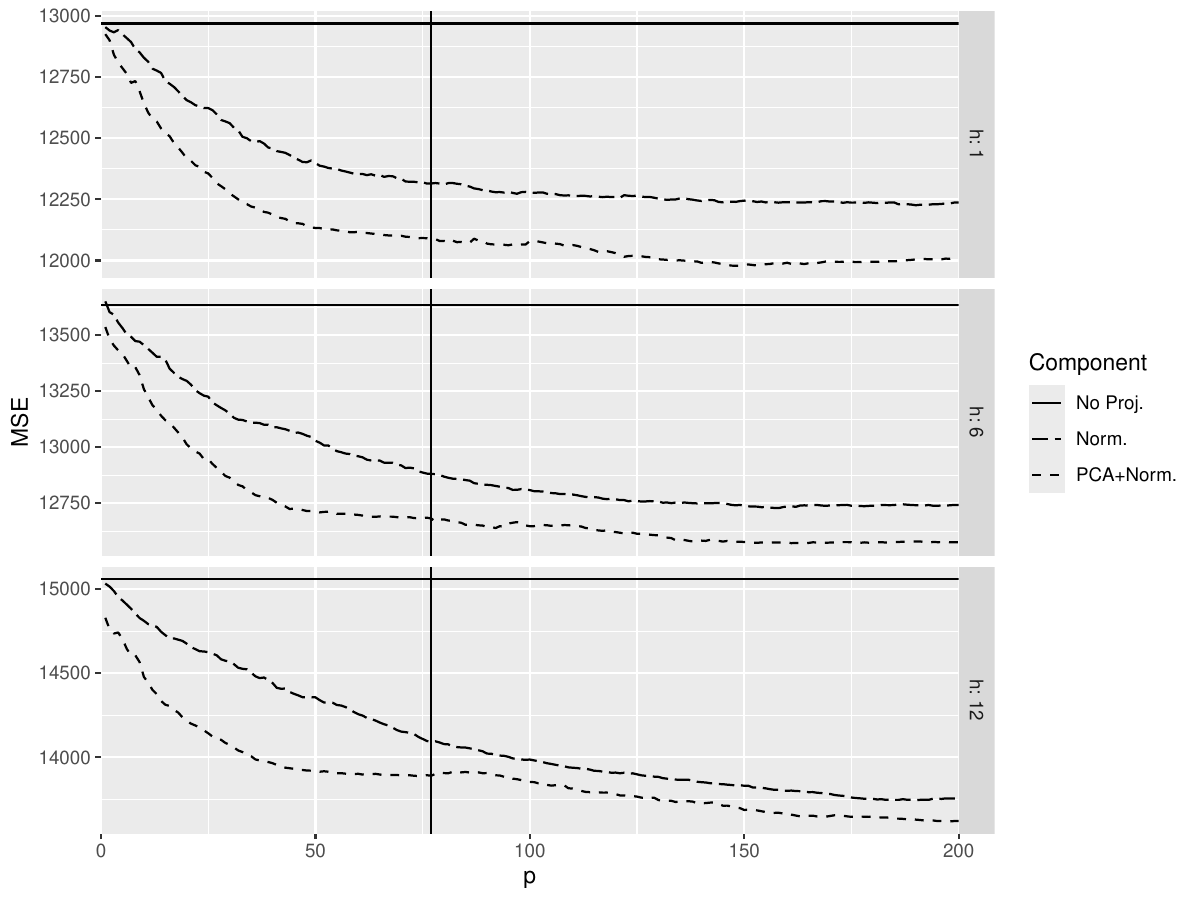}

}

\caption{\label{fig-visnights-line}Out-of-sample MSE of base and FLAP
forecasts as the number of components \(p\) increases, for forecast
horizons \(1\), \(6\) and \(12\), using the visitor nights data. The
solid horizontal lines show the MSE for the base forecasts while the
dashed lines show the MSEs of the FLAP forecasts. The vertical black
line indicates the location of \(p=m\), the number of series.}

\end{figure}%

MCB and MSE plots for \(h=1\), \(6\) and \(12\) (as described in
Section~\ref{sec-evaluation}) are presented in
Figure~\ref{fig-visnights-mcb-series} and
Figure~\ref{fig-visnights-line}, respectively. The FLAP forecasts in
general outperform the base forecasts, which is consistent with the
findings in Section~\ref{sec-simulation}. In
Figure~\ref{fig-visnights-mcb-series}, the base forecasts are always
ranked last. Interestingly, FLAP forecasts improve over base forecasts,
even when FLAP is using either one or at most two principal components.
These improvements are statistically significant across all forecast
horizons.

We highlight two findings that diverge from those observed in
Section~\ref{sec-simulation}.

First, in contrast to the results from Section~\ref{sec-simulation}, the
method implemented to construct components matters. Using principal
components in FLAP improves base forecasts more compared to using random
components with normal weights. Figure~\ref{fig-visnights-mcb-series}
shows that the best ranked forecasts are the FLAP forecasts using either
the \(p=m=77\) principal components, or complementing these with random
components so that \(p=200\). These improve on FLAP forecasts using only
\(p=200\) random components and the improvements are statistically
significant.

This is also clearly reflected in Figure~\ref{fig-visnights-line}, where
the reduction in terms of MSE from using PCA is always lower, even after
the \(p=m\) principal components are exhausted and random components
with weights generated from a normal distribution are added.

PCA aims to find components along the direction where the data varies
the most. The first principal components accounts for the maximum
variance in the data. Subsequent principal components are orthogonal to
the previous ones and capture the maximum remaining variance. Based on
this variance maximisation and ranking of PCA, we propose two potential
explanations for its superior performance.

\begin{enumerate}
\def\labelenumi{\arabic{enumi}.}
\tightlist
\item
  Optimality: By maximising and ranking the variance of PCs from largest
  to smallest, we ensure that the projection utilises components
  containing significantly more information (as measured by variance)
  compared to randomly weighted components.
\item
  Diversity: Actively seeking principal components with the highest
  variance, results in the incorporation of a more diverse set of
  components into the projection.
\end{enumerate}

Second, consistent with the results in Section~\ref{sec-simulation}, the
MSEs of FLAP forecasts shown in Figure~\ref{fig-visnights-line} decrease
as the number of components increases. However, this decreases ceases,
especially for \(h=1\), approximately for \(p>150\). This is also
reflected in the results presented in
Figure~\ref{fig-visnights-mcb-series}, where the difference in ranking
of the two best methods is not statistically significant, even though
they use very different numbers of components (\(p=m=77\) and
\(p=200\)).

Choosing the number of components is a trade-off between the increasing
estimation error as the dimension of forecast error variance
\(\bm{W}_h\) increases, and the additional benefit brought by the
information embedded in the new components. Of course this all depends
on the complexity of the DGP. For the visitor nights data, the benefit
of components over and above the estimation error, diminishes after the
number of components reaches \(p=m=77\).

\subsection{FRED-MD}\label{sec-fred-md}

The FRED-MD \autocite{McCNg2016} is a popular monthly data set of
macroeconomic variables, and shares similar properties with the
\textcite{StoWat2002} data. We downloaded and transformed the data using
the \texttt{fbi} package \autocite{fbi}. The period we use for this
exercise is from January 1959 to September 2023, containing \(777\)
observations. Following \textcite{McCNg2016}, we replace observations
that deviate from the sample median by more than 10 interquartile ranges
(which are recognised as outliers), with missing values. We then drop
any series with more than \(5\)\% observations missing. This results in
\(m=122\) series. We fill in the missing values using the
expectation-maximization (EM) algorithm described in
\textcite{StoWat2002a} with \(8\) factors. The number \(8\) is
identified by \textcite{McCNg2016}, albeit with a different time span.

In order to relate to the theoretical forecast error variance reduction,
we use MSE as the error measure, instead of other scaled or percentage
error measures. To reliably calculate MSE over series with different
scales, we demean the series and scale them to have variance \(1\). The
MSEs are calculated on this standardised scale without
back-transformation.

We evaluate the performance of forecasts using time series
cross-validation. We start with \(300\) observations in the first
training set and the following \(12\) observations as the test set. We
repeat the exercise by expanding the training set increasing by \(1\)
observation at a time. This provides us with \(466\) forecasts for each
of the forecast horizons from \(h=1\) to \(12\) for evaluation.

\begin{figure}

\centering{

\includegraphics{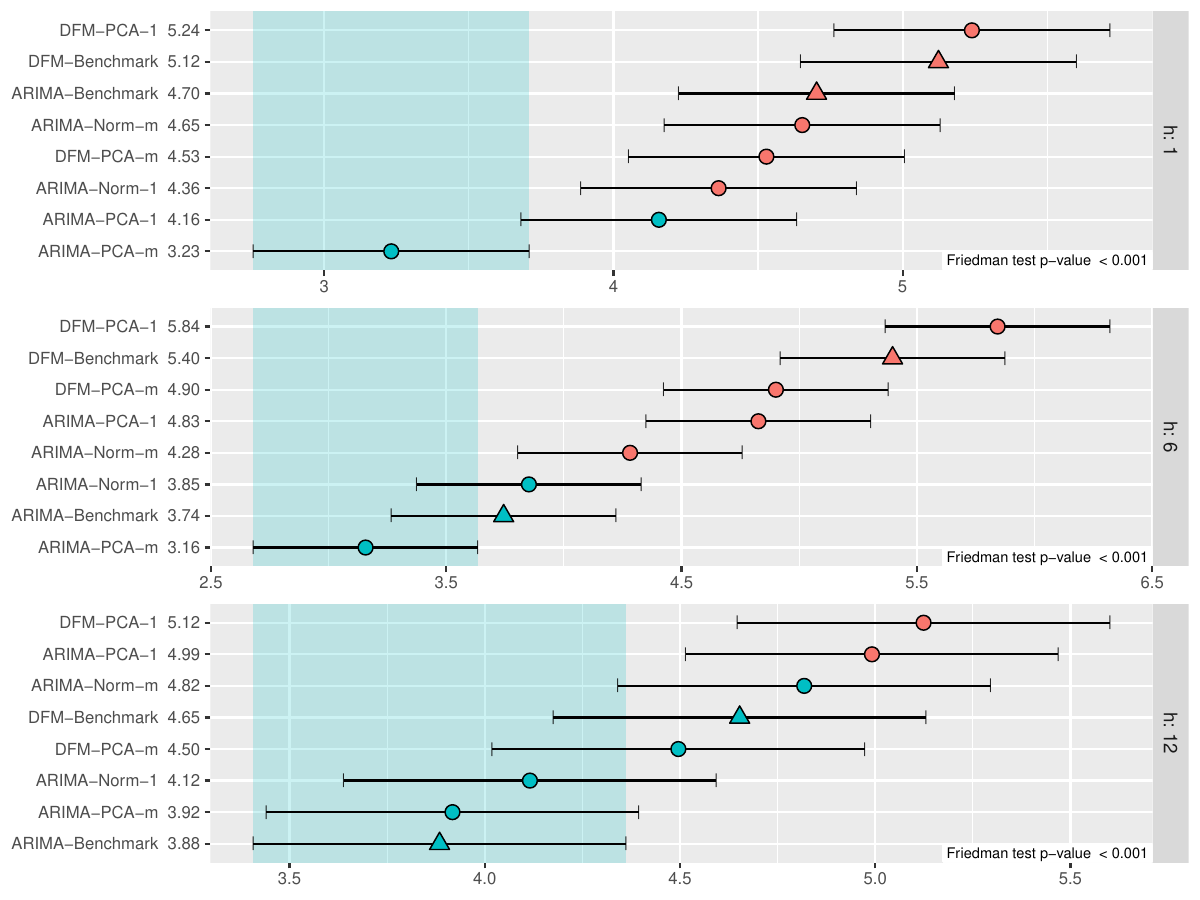}

}

\caption{\label{fig-fred-md-mcb-series}Average ranks of \(1\)-, \(6\)-
and \(12\)-step-ahead cross-validation MSE of different model and
component specifications on the FRED-MD data. The methods using forecast
projection are named as ``Model -- Component Weights -- Number of
Components''. The base models are named as ``Model -- Benchmark'' and
these points are marked with triangles. The shaded region is the
confidence interval of the best performing model. Methods outside the
shaded region are significantly worse than the best model.}

\end{figure}

\begin{figure}

\centering{

\includegraphics{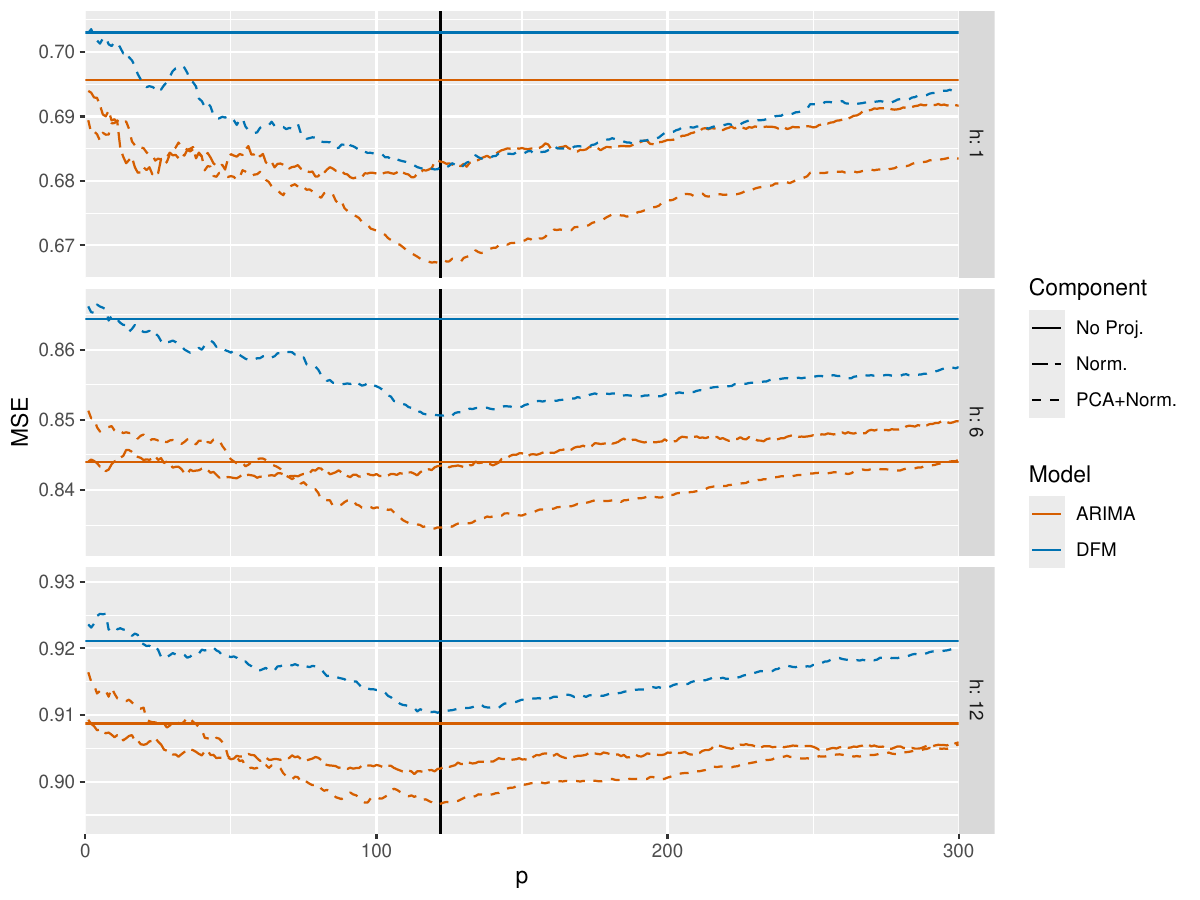}

}

\caption{\label{fig-fred-md-line}Out-of-sample MSE for base and FLAP
forecasts as the number of components \(p\) increases, for forecast
horizons \(1\), \(6\) and \(12\), using the FRED-MD data. The solid
horizontal lines show the MSE for the base forecasts while the dashed
lines show the MSEs of the FLAP forecasts. The vertical black line
indicates the location of \(p=m\) the number of series.}

\end{figure}

We generate base forecasts from ARIMA models using \texttt{auto.arima()}
function from the \texttt{forecast} package \autocite{forecast} with the
default settings, and DFMs. The ranges of the meta-parameters in the
DFMs are \(1\leq k \leq 8\) (since \(8\) factors are identified and used
to fill in the missing values), \(1 \leq n \leq 3\) and
\(0 \leq s \leq 6\). For more details see Section~\ref{sec-benchmarks}.
The MCB and MSE plots for \(h=1\), \(6\) and \(12\) are presented in
Figure~\ref{fig-fred-md-mcb-series} and Figure~\ref{fig-fred-md-line},
respectively.

The results presented in Figure~\ref{fig-fred-md-mcb-series} show that
the ARIMA base forecasts rank better that the DFM base forecasts. This
difference is statistically significant for \(h=6\). Likewise, ARIMA
base forecasts return a lower MSE compared to DFM as shown in
Figure~\ref{fig-fred-md-line} across all forecast horizons.

The best ranked forecasts across all forecast horizons are once again
FLAP forecasts generated using principal components. The improvements
over the ARIMA base forecasts are significantly significant for \(h=1\)
and \(6\). Furthermore, FLAP forecasts using PCA rank better than FLAP
forecasts using components with random weights, and these differences
are statistically significant. This is also observed in terms of MSE in
Figure~\ref{fig-fred-md-line}, reaffirming our findings based on the
tourism data set discussed in
Section~\ref{sec-australian-domestic-tourism}.
Figure~\ref{fig-fred-md-line} shows that both FLAP ARIMA and FLAP DFM
forecasts, seem to be worse than the base forecasts when using a small
number of components \(p\) for \(h=6\) and \(12\). However, this
improves as \(p\) increases and FLAP forecasts outperform the base
forecasts gradually. The trade-off between the benefit of including more
components and the loss in degrees of freedom due to estimation in a
large dimension is once again observed. However in this case it seems to
be more extreme, compared to the tourism example, with MSEs starting to
increase for \(p>m\).

As we have seen in Section~\ref{sec-simulation} and
Section~\ref{sec-australian-domestic-tourism}, \(m\) is not always a
clear cut-off point in determining the number of components to use. The
performance of FLAP is jointly determined by the number of series \(m\),
the sample size \(T\), the component construction method, and the DGP.
The example of FRED-MD shows the importance of PCA, as \(m\) is the
point that the component changes from PCA to random normal weighted
linear combinations, implying that PCA can exploit the information in
the data while random weights cannot. This is more obvious for \(h=1\)
and \(h=6\), as PCA works when \(p<m\), but random normal weights do not
seem to work from the beginning.

\section{Conclusion}\label{sec-conclusion}

The proposed forecast linear augmented projection (FLAP) method has been
shown to be a simple but effective way to reduce forecast error variance
of any multivariate forecasting problem. It simply involves augmenting
the data with linear combinations, forecasting these and then projecting
the augmented vector of forecasts. We have shown theoretically that FLAP
will continue to reduce forecast error covariance as more components are
added, assuming that the forecast error covariance matrix is known. In
practice, a plug-in estimate of this covariance matrix can be used, and
in both simulated and empirical data we demonstrate that a simple
shrinkage estimator does indeed lead to improvements in forecast
accuracy. Regarding the construction of components, we find that PCA in
practice achieves significant improvements in forecast accuracy. Another
appealing property of FLAP is that the projection step can even
compensate for a poor choice of base forecasting model. This is
particularly attractive since it makes FLAP robust against model
misspecification in the base forecasting step.

One outstanding issue is to find alternatives to PCA to select component
weights. For example, \textcite{Goerg2013-dg} proposed ``forecastable
components'' that are optimal in the sense of minimising the forecast
error variance of the components, while \textcite{Matteson2011-jf}
proposed ``dynamic orthogonal components'' that reduce a multivariate
time series to a set of uncorrelated univariate time series. It would be
interesting to explore whether these components (or other similar
suggestions) can be used effectively in FLAP. Another route to improving
FLAP may be found by optimizing Equation~\ref{eq-obj} over \(\bm{\Phi}\)
and \(\bm{G}\) rather than just \(\bm{G}\).

Component construction should be studied together with the selection of
the forecast model since both the weight matrix \(\bm{\Phi}\) and the
base forecast error variance \(\bm{W}_h\) can affect the projection
simultaneously in Equation~\ref{eq-M}. This is likely to be related to
forecast combination methods, focusing on the properties of the base
forecasts, and the diversity and robustness of the forecast model and
components. Examples of studies on this issue in the forecast
combination literature include \textcite{BatDua1995},
\textcite{KanEtAl2022} and \textcite{LicWin2020}.

Finally, while FLAP is motivated by the forecast reconciliation
literature, our focus here is very much on multivariate time series with
no constraints. However, it would be possible to use both forecast
reconciliation and forecast projection together. This may be
particularly useful when there are relationships between series that are
not captured in the known hierarchical structure.

\section*{Acknowledgements}\label{acknowledgements}
\addcontentsline{toc}{section}{Acknowledgements}

We thank Daniele Girolimetto for contributing to the initial proof of
Theorem~\ref{thm-monotone} in his unpublished work.

\printbibliography[title=References]

\end{document}